\documentclass[superscriptaddress,aps,prl,reprint]{revtex4-2} 
\usepackage[english]{babel}
\usepackage{amsthm}

\usepackage{booktabs}
\usepackage{float} 
\usepackage{lipsum}
\usepackage{graphicx,subfigure,color}
\usepackage[caption=false]{subfig}
\usepackage{amssymb}
\usepackage{amsmath}
\usepackage{commath}
\usepackage{graphicx,bm}
\usepackage{verbatim}
\usepackage{graphicx}
\usepackage{dcolumn}
\usepackage{bm}

\usepackage{xcolor}
 
\newtheorem{thm}{Theorem}

\newcommand{\nn}{\nonumber}

 \newcommand{\w}{\omega}

\newcommand{\be}{\begin{equation}}
	\newcommand{\ee}{\end{equation}} 
\newcommand{\ba}{\begin{array}}
	\newcommand{\ea}{\end{array}}
\newcommand{\bea}{\begin{eqnarray}}
	\newcommand{\eea}{\end{eqnarray}}

\newcommand{\half}{\frac{1}{2}}

\usepackage{multirow}
\begin{document} 
	\preprint{APS/123-QED}	
	\title{  Quantum Scaling  Dimension from the  Equivalence principle  }
	\author{Yoon-Seok Choun}
	\email{ychoun@gmail.com}
	\affiliation{ Department of Physics, Hanyang University, Seoul 04763, South Korea
	}
	\author{Ki-Seok Kim}
	\email{tkfkd@postech.ac.kr}
	\affiliation{Department of Physics, POSTECH, Pohang, Gyeongbuk 37673, Korea}%
	\affiliation{Asia Pacific Center for Theoretical Physics (APCTP), Pohang, Gyeongbuk 37673, Korea}
	\author{Sang-Jin Sin$^{1,}$}%
	\email{sangjin.sin@gmail.com}
	\author{Taewon Yuk$^{1,}$}%
	\email{tae1yuk@gmail.com}

	\date{\today}
	
	\begin{abstract} 
		We propose a method to constrain the scaling dimension 
		of the operators of the strongly interacting systems (SIS) using the holographic setup. 
		We demonstrate our method  using the holographic superconductor theory.  The idea is to consider  the inside   as well as  the outside of the AdS black hole in which the gap equations has higher order singularities. Then the equivalence principle requests the solution  be smoothly connected at the horizon, which request the vanishing of log divergent term as well as an indefinite conditionally convergent  terms that can lead to any real number according to Riemann. 
		As a result,  one gets quantized values of the scaling dimension of the condensing  operator. This is a pleasant surprise because  so far one gets  the constraints on the scaling dimension  only by a hard  analysis with  bootstrap ansatz. 	
		%
		
	\end{abstract}
	
	\maketitle
	
	\paragraph*{Introduction:} 
	Recently, it has been recognized that the inside of the black hole should play an important role to understand entanglement  and the  information   puzzle of   black hole\cite{Maldacena:2013xja,Penington:2019kki,Almheiri:2019qdq}, hence it should also have deep implication to  the general  property of the quantum field theory and possibly on the  quantum material  at the finite temperature  through the gauge gravity duality.  
	
	In fact,  constraining a physical parameter in quantum field theory is  very hard task in general. 
	One exceptional example  so far is the   bootstrap program \cite{RevModPhys.91.015002}  to constrain the operator scaling dimension. Obviously, the analysis is highly  non-trivial so that only  numerical bootstrap program works.   Gauge gravity duality \cite{Maldacena:1997re, Witten:1998qj,Gubser:1998bc} in principle might provide an opportunity to constrain the  anomalous dimension  in terms of the classical geometry. However,   the scaling dimension in  gauge gravity duality has been assumed to be  arbitrary so that one  can set $\Delta$  by  hand.  \\
	
	Recently, we developed a  method to constrain the scaling dimension of the operator   in a few examples using the holographic set up. 
	In 	 \cite{Choun:2021pvs},   we  showed  that  the scaling dimension  of the cooper pair operator is restricted to $\Delta>1$  studying  the gap equation   of the  holographic superconductors  \cite{gubser2008gravity,Hart2008,gubser2008gravity,benini2011holographic} 	near the horizon.  
	In  \cite{choun2022quantized},   we requested that the solution must be extended to the inside of the black hole following the work of  Horowits et.al\cite{Hartnoll:2020fhc}. 
	To get the regular solution everywhere, we demanded that the solution be of polynomial type after factoring out the near- singularity-behavior. As a result, only quantized pairs  of coupling  and  scaling dimension are allowed. It works for  3+1 dimensional system  but not for 2+1 dimensional one.  \\

	In this letter, we report an  unified  method which works 
	for 2+1 dimensional system  as well as  for 3+1. 
	Our analysis shows that there are only three possible values of scaling dimensions: $\Delta=1.50,\; 2.88,\; 3.00 $ in the regime $-9/4<m^2<0$.  
	We will show that if we request the solutions  inside and outside of the horizon are connected smoothly with   well-defined single value  at every regular  points, then only discrete values of scaling dimensions are allowed.  
	The phenomena is an analogue of the quantization of the energy for a confined system in quantum mechanics.   The gap equation in the Reisner-Nordstrom  black hole has higher  sigularities and request more stringent  regularity conditions, which is the origin of the quantization of the parameters. 
	
	For the fixed charge density and coupling,  the scaling dimension  can have only quantized value even when we consider the outside solution only.   
	We also show that when this method is applied to the 3+1 dimensional system it reproduces the previous result.  
	Our present method works  for arbitrary tensor operator. 
	
	We will see that the horizon regularity gives the solution which is also regular at the curvature singularity as well, so that 	the inner solution should be regular everywhere inside of the black hole.  The previous authors have not seen this phenomena due to the subtlety in treating singular differential equations: the series solutions depends on the coordinate system one use. All such solutions can be considered as the analytic continuation of one another. However, they are different solutions in the sense that they have different radius of convergence regime, although not independent to one another.  Only when we use the   solution whose coordinate center is a regular point, our claimed result  can be derived easily. The   analysis in ref. \cite{Hartnoll:2020fhc} use the coordinate whose center is  either the   black hole center or the horizon, both of which are singular points so that log solutions are explicitly involved,  which makes the treatments subtle. \\
	
	\paragraph*{Set up:} 
	We consider  Abelian-Higgs model   in AdS bulk \cite{Hart2008}  
	\begin{footnotesize}
		\bea  {
			S=\int d^{d+1}x \sqrt{-|g|}\left( -\frac14 F_{\mu\nu}^{2}- |D_{\mu}\Psi |^{2}-m^{2}|\Psi|^{2} \right),} \\
		\hbox{with }  |g|=\det g_{ij}, \quad    D_{\mu}\Psi=\partial_{\mu}-igA_{\mu}, \quad F=dA, A=\Phi dt. 			\nn 
		\eea
	\end{footnotesize}
	We  use   AdS$_{d+1}$-Schwarzschild black hole  background,   
	\begin{footnotesize}
		\[ ds^2= -f(r)dt^2+\frac{dr^2}{f(r)}+r^2d\vec{x}^2, \hbox{ with }  f(r)=r^2\left(1-\frac{r_h^d}{r^d}\right),\]
	\end{footnotesize}
	where 	
	$r_h$ is the radius of the horizon.  
	The Maxwell-scalar field equations become  
	\begin{footnotesize}
		\bea
		&&  \frac{d^2 \Psi }{d z^2} -\frac{d-1+z^d}{z(1-z^d)}\frac{d \Psi }{d z}+\left[ \frac{g^2 \Phi^2}{r_h^2(1-z^d)^2}-\frac{m^2}{z^2(1-z^d)}\right]\Psi =0, \nonumber \\
		&& \frac{d^2 \Phi}{d z^2}-  \frac{d-3}{z}\frac{d \Phi}{d z } -\frac{2g^2\Psi^2}{z^2(1-z^d)}\Phi =0,
		\label{eq:3}
		\eea
	\end{footnotesize}
	with the coordinate $z= r_h/r$  with setting $r_h=1$. Therefore the regions $z>1$ and $0<z<1$ are inside and outside of the black hole respectively. We call this pair of equations   the {\it gap equations}, because  they determine the gap.  
	For even $d$, the differential equation can be written 
	in terms  of the variable $x=z^2$ so that the order of the singularity can be halved, which is the reason why the analysis in AdS$_5$ in the previous paper \cite{choun2022quantized} is far more simpler than that of the AdS$_4$. 
	
	The   boundary ($z=0$) behavior of the fields are  
	\bea \Psi(z) &=& z^{\Delta_{\pm}}\Psi^{(\pm)}(z), \;  \\ 
	\Phi(z) &=& \mu- \frac{\rho}{r_h^{d-2}}z^{d-2} +\cdots. \eea 
	Here, 	$\Delta_{\pm}=\frac{d}{2}\pm \sqrt{\frac{d^2}{4}+m^2}$
	and  $\mu$ and  $\rho$  are the chemical potential  and the charge density, respectively. 
	We limit our interests to near critical temperatures where 	probe solution can be  trusted \cite{Horo2009,Horo2011}.  
	
	\paragraph{Near  $T_c$  outside the black hole:}\label{gaga}
	We divide the domain of the solution   into   outside  and  inside   the black hole, and then request that  the solution be smoothly  connected   at the event horizon at $z=1$. 
	Right at the critical temperature $T_c$, the condensation is  zero and  $\Psi =0$, so the gap equations  becomes  decoupled and  linear: the second equation of \eqref{eq:3}  lead us to 
	$\Phi(z)= \tilde{\lambda}  r_c (1-z)$ whose insertion to the first equation of    \eqref{eq:3}
	gives the {\it decoupled linear } differential equation of $\Psi$ :  
	\begin{footnotesize}
		\begin{equation}
			-\frac{d^2 \Psi }{d z^2} +\frac{2+z^3}{z(1-z^3)}\frac{d \Psi }{d z}+  \frac{m^2}{z^2(1-z^3)} \Psi =\frac{\lambda^2}{(z^2+z+1)^2}\Psi, 
			\label{eq:10}
		\end{equation}
	\end{footnotesize}
	with $\lambda = g\tilde{\lambda}$ and  $\tilde{\lambda} =\frac{\rho}{r_c^2}$. 
	Factoring out the behavior near the boundary $z=0$ and those near  $z=\w$ and $\w^2$ with $\w=\frac{ -1+ i \sqrt{3} }{2} $,  the solution outside the horizon can be written as 
	\begin{footnotesize}
		\begin{equation}
			\Psi_{out}(z)= F(z) Y_{out}(z),
			\label{mm:1}
		\end{equation} 
		with  $F(z)=\frac{\left< \mathcal{O}_{\Delta}\right>}{\sqrt{2}r_h^{\Delta}}z^{\Delta} 	(z^2+z+1)^{-\lambda/\sqrt{3}}$.
	\end{footnotesize}   
	The equation for  $Y_{out}(z)$ is a generalized Heun's equation that has five regular singular points at $z=0,1, \w,\w^2, \infty$.
	Expanding   $ Y_{out}(z)= \sum_{n=0}^{\infty } d_n z^{n}$ and insert it to the differential equation,   we obtain the four term  recurrence relation,
	\begin{equation}
		d_{n+1}= A_n \;d_n + B_n \;d_{n-1}+C_n\;d_{n-2} ,  \hbox{  for  } n \geq 2.  \label{mm:3}
	\end{equation}
	Here, $A_n, B_n, C_n $ consists with $\Delta$,  $\lambda$ and index $n$.
	And the first three  $d_{n}$'s are  $d_{0}=1$, $d_1=A_{0}d_0$ and  
	$d_{2}=A_{1}d_1 +B_{1}d_{0}$.

	Now,  $d_n$ has two   independent solutions $	d_{n}^{(1)}$,  $	d_{n}^{(2)}$
	\cite{elaydi1996introduction},  such that 
	$
	d_n   \sim  a_0	d_{n}^{(1)} + a_1	d_{n}^{(2)} $ for large $n$ where
	\be
	d_{n}^{(1)}=  n^{-1} ,  \quad 
	d_{n}^{(2)}= \frac{    \cos \left(\frac{2 \pi  n}{3}\right)}{n^{\frac{2 \lambda }{\sqrt{3}}+1}} .\ee
	
	Here, $a_0, a_1$ consist of $\Delta$ and $\lambda$. 
	A harmonic series   $\sum_{n}^{\infty}d_n^{(1)} z^n$ develops $a_{0} \log(1-z)$ behavior near horizon. 
	For the horizon regularity, we require 
	\be a_{0}(\lambda,\Delta)=0, \ee  
	which is enough to make the solution   convergent at  the horizon,  $z=1$.  
	As a result we have a relation between  the scaling dimension and coupling, 
	which is  given by the  red curves in Fig.~\ref{intersection1} (a)  by numerical analysis.    
	
	\begin{figure}[!htb]
		\centering
		\subfigure[]
		{ \includegraphics[width=0.36\linewidth]{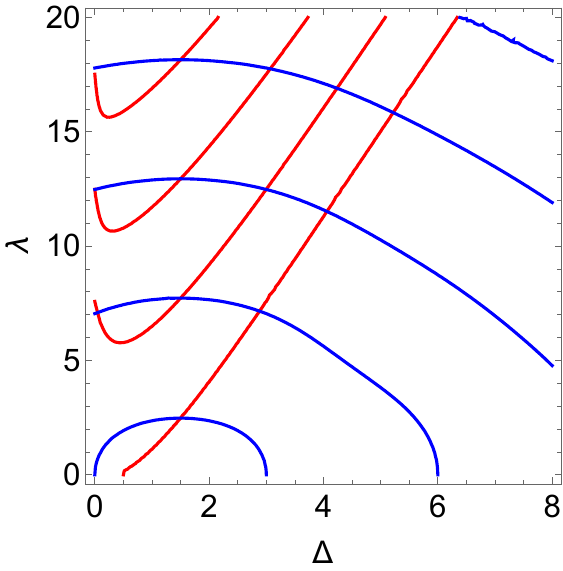}}  
		\subfigure[]
		{ \includegraphics[width=0.6\linewidth]{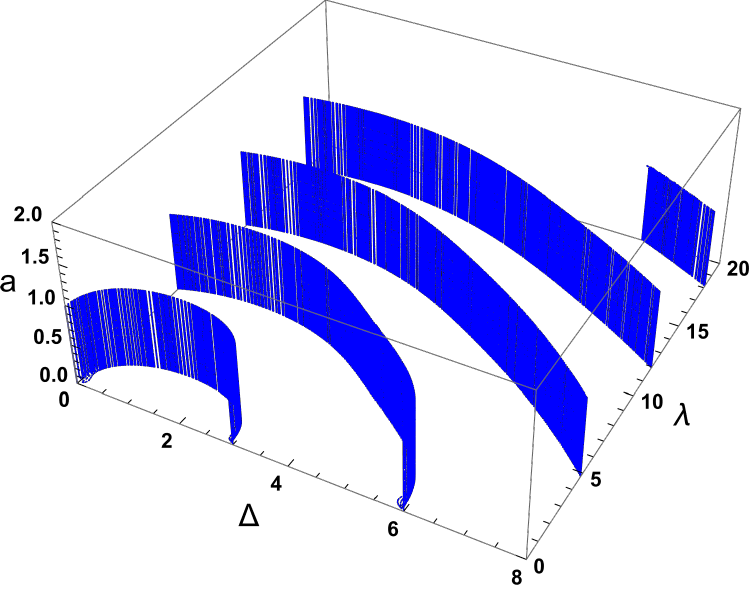}} 
		\caption{  (a) Conditions for the horizon regularity: 	Red and Blue curves are   those for outside and inside solutions respectively. 
			Smooth  connecting   them   at the horizon restrict  the  intersection points as the only allowed values.   (b)  \small 
			In all coordinate $\zeta_a=a-\frac{1}{z}$ with  $0 \leq a \leq 1$, $ \det \mbox{M}(\Delta, \lambda, a)=0 $ for an inside solution gives us the horizon regularity condition with well-definedness. When projected, the curve exactly coincides with the blue curves in (a).    } 
		\label{intersection1}
	\end{figure}  
	
	\paragraph*{Solution inside.}
	We now consider the solution inside the black hole where it can have  a $\log$ singularity.  We will show, however,  that the horizon regularity   implies the   regularity at the center of black hole as well, which is a key observation of this paper. 
	Due to the log divergence that appears here and there in the original variable $r=1/z$, we analytically continue the solution using the  variable change and demonstrate the complex analytic structure of the solution of the differential equation  in the nearby variable where there is not log in the second solution.  We will then come back to the original coordinate variable. 
	
	For this purpose, 
	we   introduce  one parameter family  of  coordinates 
	\be 
	\zeta_a=a-\frac{1}{z}, 
	\label{gen}
	\ee
	centered at $r=a$  with    $0\leq a \leq1$.  We will consider $a$ as real but $z, r, \zeta_a$ as complex variables.  
	Below we will see that the analysis does not depend on the 
	variable one use and we 
	divide the regime of $a$  into two: (1) $0\leq a \leq 1/2$ where our analysis works straightforwardly and (2) $1/2 < a \leq 1$ where it  works with subtlety.  
	To deliver the general idea, 	 we will examine   four cases   $a= 0,  1/3, 1/2, 2/3, 1$ separately and explain   why $\Delta$ is quantized. After we demonstrate the general structure of the analysis, we come back to the original case where $a=0$. 
	\begin{figure}[!htb]
		\centering
		\subfigure[$ a=1/2$]
		{ \includegraphics[width=0.4\linewidth]{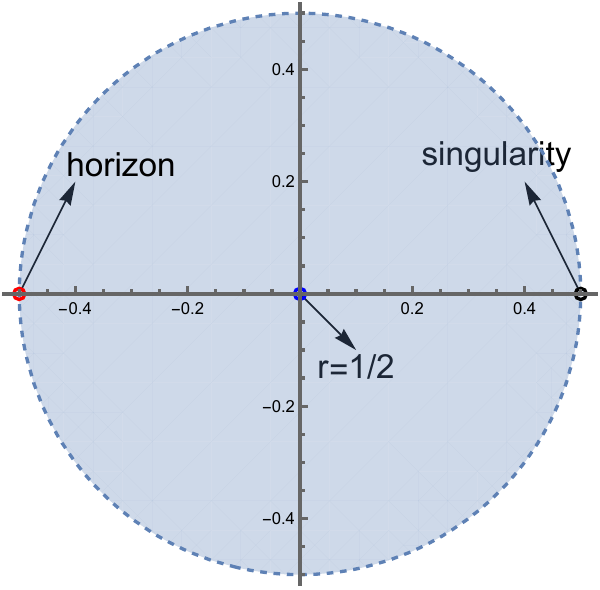}} 
		\subfigure[$  a=1/3$]	{ \includegraphics[width=0.4\linewidth]{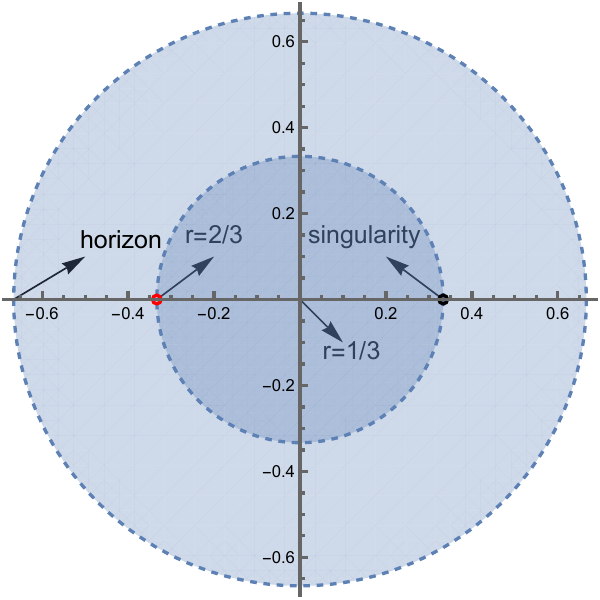}}  
		\caption{ Two working coordinates  $\zeta_a$ for $a=1/2$ and $\zeta_a$ for $a=1/3$.   } 
		\label{domain}
	\end{figure}  
	\paragraph*{$\bullet$  General $a$:} 
	
	Factoring out the behavior near $z=0$, we have 
	\begin{footnotesize}
		\begin{equation}
			\Psi_{in}(\zeta_a)= \mathcal{A} \Big[ F( {z})  (1/z)^{\Delta -\frac{2\lambda}{\sqrt{3}}}\Big]_{z= z(\zeta_a) }y_{in}(\zeta_a). 
			\label{mm:14}
		\end{equation}   
	\end{footnotesize}
	Here, $\mathcal{A}$ is a normalization constant determined by the continuity of the inside and outside solutions.  
	Under the  substitution $\zeta_a =a- 1/z:=z(\zeta_a)$,    the number of  singularties 
	of eq. (5)   is invariant.  
	The equation for  	$y_{in}(\zeta_a)$ is a Fuchsian  differential equation that has five regular singular points at $\zeta_a=a, a-1, \frac{1}{2}  (2 a \pm i \sqrt{3}+1 ), \infty$. 
	Inserting $ y_{in}(\zeta_a)= \sum_{n=0}^{\infty } e_n \zeta_a^{n}$  to the differential equation,   we obtain the five term  recurrence relation: 
	\begin{equation} 
		e_{n+2}=	A_n e_{n+1}+B_n e_n+C_n e_{n-1}+D_n e_{n-2} .  \label{mm:200}
	\end{equation} 
	Here  we define $e_n=0$ for negative $n$. 
	Here, $A_n, B_n, C_n, D_n $  depend on  $\Delta$,  $\lambda$, $a$ and $n$.  
	The eq. 	\eqref{mm:200} can be represented in the matrix form as
	\begin{equation}
		\sum_{m=0}^{\infty} \mbox{M}_{n,m}(\Delta, \lambda, a) \; e_m =0,  \quad  (n=0, 1, 2, \cdots). \label{mmr:26}
	\end{equation} 
	The basic idea is that to have the non-trivial set of $e_n$, the determinant of the matrix $M_{m,n}$ should be zero. 
	To handle this infinite dimensional matrix numerically, we first truncate to $n=N+2$ for large enough $N$. 
	If we can set $e_{N+1}=e_{N+2}=0$,  $\mbox{M} $ becomes $N \times N$ matrix.	Only when $\det \mbox{M} = 0$ is satisfied,  
	the homogeneous equation \eqref{mmr:26} allows a nontrivial solution.    The converse is also true, because, with truncation, if $ \det \mbox{M}_{N\times N}  = 0$, then $e_{N+1}=0, \;  e_{N+2}=0$.
	Therefore, 
	\be 
	e_{N+1}=0, \;  e_{N+2}=0  \leftrightarrow  \det \mbox{M}_{N\times N} = 0.
	\label{determinant}
	\ee 
	Now,  for large $n$, $e_n$  can be written as sum of three terms  $	e_{n}^{(1)}$,  $	e_{n}^{(2)}$,  $e_{n}^{(3)}$
	\cite{elaydi1996introduction}: 
	$  e_n  \sim  \nu_0	e_{n}^{(1)} + \nu_1	e_{n}^{(2)}+ \nu_2	e_{n}^{(3)} $ where
	\bea 
	e_{n}^{(1)} &=&   {\left(\frac{1}{a-1}\right)^n}/{n}, \quad e_{n}^{(2)}=  {\left(\frac{1}{a}\right)^n}/{n}, \\
	e_{n}^{(3)}&=&  {\left(\frac{1}{a^2+a+1}\right)^{n/2} \cos (\alpha  n)}/{n^{\frac{2 \lambda }{\sqrt{3}}+1}} \eea 
	with 
	$\alpha =\tan ^{-1}\left(\frac{\sqrt{3}}{2 a+1}\right)$. Here, $\nu_0, \nu_1, \nu_2$ consist of $\Delta$, $\lambda$ and $a$.
	On the other hand, 
	$e_n$ can be  written as a linear combination of $e_0$ and $e_1$: 
	\be 
	e_n  = e_0	V_n(\Delta,\lambda,a)   + e_1 W_n(\Delta,\lambda,a),
	\label{rrr:30}
	\ee
	because   all the rest of $e_n$ are given as   polynomials of  $A_n,B_n,...$ and  it is linear in $e_0,e_1$.  Then  
	\begin{footnotesize}
		\be 
		\begin{pmatrix} 	 e_{N+1}  \\ e_{N+2} \end{pmatrix}=0  \Leftrightarrow
		\begin{vmatrix} 	  
			V_{N+1} &  W_{N+1}\\ V_{N+2 }& W_{N+2}  \end{vmatrix}=0 . 
		\label{rrr:40}
		\ee
	\end{footnotesize}
	Since   $\nu_2	e_{n}^{(3)} $  is negligible for    sufficiently large $n$,   
	\be    \lim_{n\to \infty}e_n = \nu_0	e_{n}^{(1)} + \nu_1	e_{n}^{(2)}.
	\ee 
	Then 	in the limit  $n \to \infty$, $\nu_0$ and $\nu_1$ are  given  by 
	\begin{footnotesize}
		\begin{eqnarray}
			\nu _0 &=&   \frac{(-1)^n \left((n+2) e_{n+2}-2 (n+1) e_{n+1}\right)}{2^{n+3}} \nonumber\\
			\nu _1 &=& \frac{2 (n+1) e_{n+1}+(n+2) e_{n+2}}{2^{n+3}}
			\label{rrr:50} 
		\end{eqnarray}
	\end{footnotesize}
	Substitute \eqref{rrr:30} into  \eqref{rrr:50}, we see 
	\begin{footnotesize}
		\be 
		\nu_0=\nu_1=0  \leftrightarrow V_{n+2} W_{n+1}-V_{n+1} W_{n+2}=0. 
		\label{rrr:60}
		\ee
	\end{footnotesize}
	From  \eqref{determinant},  \eqref{rrr:40},\eqref{rrr:60},  we have 
	\begin{footnotesize}
		\be 
		e_{N+1}= e_{N+2}=0 \leftrightarrow	\nu_0=0, \;  \nu_1=0  \leftrightarrow  \det \mbox{M}  = 0.
		\label{rrr:61}
		\ee
	\end{footnotesize} 
	Now we analyze the convergence of the series solution at the horizon which is located at $\zeta_a=a-1$: 
	\be y_{in}(\zeta_a=a-1) \sim 
	\nu_0 \sum _{n }^{\infty } \frac{1}{n}+  \nu_1 \sum _{n }^{\infty } (1-1/a)^n  /{n}.\ee 
	Depending on the $a$'s regime $a\in(0,1/2]$ or $a\in( 1/2,1)$, the analysis is different.  To avoid possible confusion, we take   typical points in each  regime and pay special attentian to the boundary values of the regime, $a=0,1/2,1$.

	\paragraph*{$\bullet$  $\;a=1/2$, the simplest   case:} 
	First,  consider  $a=1/2$.   
	In Fig.~\ref{domain}(a),  the  left red dot  is the horizon and the  right  dot  is the position of the black hole center.   These two points are located at the boundary of the convergence domain for the coordinate with $a=1/2$, which is $|\zeta_a|<1/2$.   
	At the horizon, 
	the asymptotic series  is 
	\be y_{in}(\zeta_a=-\half) \sim 
	\nu_0 \sum _{n }^{\infty } \frac{1}{n}+  \nu_1 \sum _{n }^{\infty } \frac{\left(-1 \right)^n}{n}.\ee 
	For the regularity at the horizon, we have to require $\nu_0 =0$, since the harmonic series $\sum _{n }^{\infty } \frac{1}{n}$ has $\log$ divergence.  Furthermore the sum  $\sum _{n }^{\infty } \frac{\left(-1 \right)^n}{n}$,  according to Riemann\cite{riemann1867ueber,kline1990mathematical}, 
	may be rearranged to converge to any prescribed real value. 
	Therefore, for the well-definedness of the series solution, we have to set 
	\be 
	\nu_1 =0,  \hbox{ as well as } \nu_0=0. \label{regularity}\ee 
	Now,  at the black hole singularity,  the solution's asymptotic piece  becomes 
	\be 
	y_{in}(\zeta_a=\half)  \sim  \nu_0 \sum _{n }^{\infty } \frac{\left(-1 \right)^n}{n}+  \nu_1 \sum _{n }^{\infty } \frac{1}{n}, \ee 
	which vanishes by \eqref{regularity}. 
	This means that the horizon regularity implies that  at the center of black hole as promised. 
	
	In practical calculation, it is better to calculate $\det \mbox{M}  =0$ 
	instead of 	$\nu_0=\nu_1=0$.   
	Blue curves in Fig.~\ref{intersection1}(a)    are solutions of  $\det \mbox{M}  =0$ for BH inside.  
	The intersection points of red and blue curves  give the quantized value of $\lambda$ and $\Delta$.   
	To list the first a few of them,  
	$$ (\Delta,\lambda)=(3/2,2.48), (3/2,7.72), (3/2,12.94),  \cdots, $$ 
	which are all lying on the line $\Delta=3/2$. 
	For   $\Delta< 3$    there are only two  more intersection points    $ (\Delta,\lambda)=(2.88, 7.16),\; (3.00,12.48)$ other than the $\Delta=3/2$ series. \\
	
	%
	\paragraph*{$\bullet$  $\;a=1/3$, a  generic      case  of $a\in(0,1/2)$:}\label{horowitz1} 
	Let's now go to the coordinate  
	$\zeta_a= \frac{1}{3}-\frac{1}{z}$     centered at $r=1/3$.  
	See  Fig.~\ref{domain}(b). 
	The red point is $r=2/3$, the left boundary of the convergence, which belongs to the  inside of a black hole. The  black dot  is the position of the black hole singularity. 	

	For  the regularity at horizon where $\eta=-2/3$, we have to require $\nu _0=0, \nu _1=0$. 
	Therefore, 
	\be 
	\nu _0=0, \nu _1=0  \leftrightarrow \det \mbox{M}'   = 0.
	\label{determinant1}
	\ee 
	Then the solution at the singularity is   regular because when $  \nu _0=0, \nu _1 =0$, the  solution's value at $r=0$ is 
	$$y_{in}(\zeta_a)|_{\zeta_a=1/3}=\nu_0 \sum(-\half)^n +\nu_1 \sum  \frac{1}{n} + \nu_2 \sum   \frac{e_{n}^{(3)}}{3^n},$$ 
	which is finite. 
	
	Our numerical calculation showed that the locus of $\det \mbox{M}' = 0$ in $(\Delta,\lambda)$ space  is completely overlapping with that of $\det \mbox{M} = 0$.  

	\paragraph*{ $\bullet$  $a=2/3$, a generic   case for  $a\in (1/2,1)$ with subtlety:}\label{horowitz2}  
	
	Now we consider the coordinate 
	whose center is closer to the horizon than the singularity. We take for example $\zeta_a= \frac{2}{3}-\frac{1}{z}$. Its    convergence regime is  $|\zeta_a|<\frac{1}{3} $ by the ratio test.  
	Its  origin  $\zeta_a=0$  is at $r=2/3$.   
	The red point is the horizon. The black one is $r=1/3$ which is    inside the black hole.  	 See Fig.~\ref{domain1}(a). 
	Here, imposing the regularity at the horizon, $\zeta_a=-1/3$, 
	expands  the convergence regime to $|\zeta_a|<2/3$.  Now, if we    impose  	the well-definedness at the boundary of the convergence,   $\zeta_a=-2/3$ or $r=4/3$, it   again  imply  the regularity at the black hole singularity which is located at $\zeta_a=2/3$.   Furthermore it gives the matrix equation $\det M''=0$ whose locus in $\Delta$-$\lambda$ space is again turns out to be identical to that of $\det M=0$ which we met in $a=1/2$ and $a=1/3$ cases.  See the supplementary material for this detail. 
	While this result is   sensible, one might  say  that  since $\zeta_a=-2/3$  or $r=4/3$  is  located outside  the black hole while we are considering inside solution, therefore there is no physical reason to impose the regularity there. \\
	
	All coordinate $\zeta_a$ with $1/2 < a\leq 1$ has such subtlety.  
	Such   apparent  subtlety can  be resolved  in two steps: first,  notice 
	that the different coordinates  describe   analytically continued function  from the original solution and they are defined in different region so that  $y_{in}(\zeta_a)$  describes  partially outside  region  as well as the inside region.   Then   we should agree  that 
	any   solution should be regular at a point  as far as the point in question is not the black hole curvature singularity.     
	Our observation  is that although we can not request the regularity at the black hole singularity,  the regularlity requested at somewhere else automatically garantees the regularity at the signularity. 
	It turns out that,  regardless of the coordinate we choose, the regularity condition 
	turns out to give the identical locus in $\Delta$-$\lambda$ plane, which is rather remarkabe. 
	
	\begin{figure}[!htb]
		\centering
		\subfigure[$  a=2/3$]
		{ \includegraphics[width=0.4\linewidth]{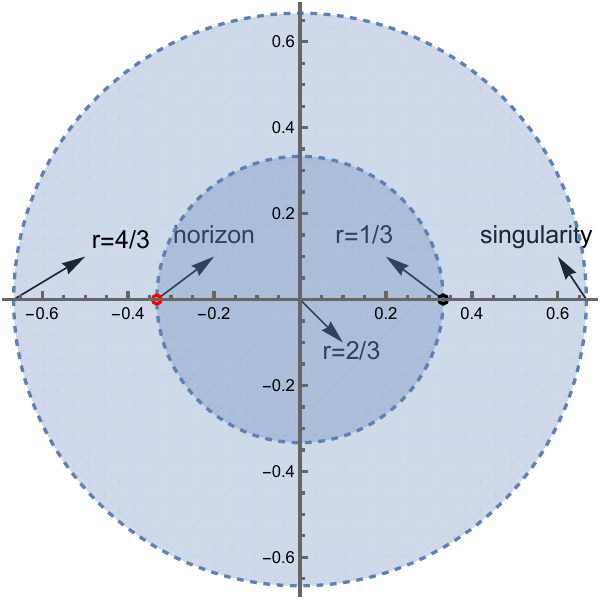}}  
		\subfigure[$ a=1$]
		{ \includegraphics[width=0.4\linewidth]{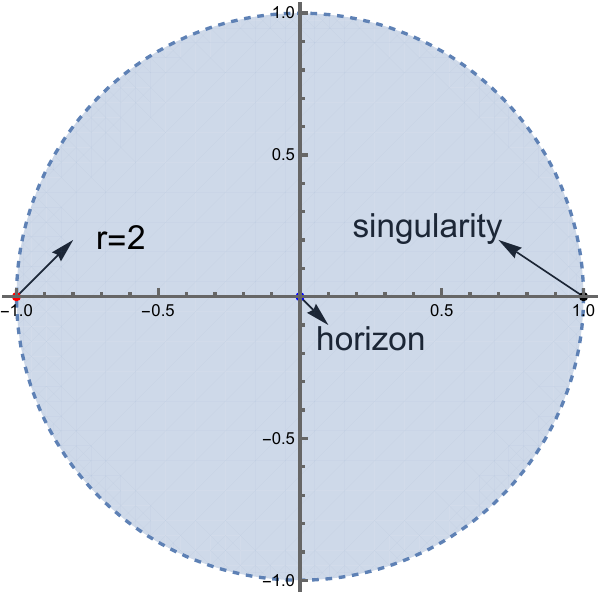}} 
		\caption{  Two non-working coordinates.   } 
		\label{domain1}
	\end{figure}  
	
	\paragraph*{ $\bullet$ $a=1$: the horizon centered coordinate:}\label{horowitz3}  
	
	We now take   the coordinate $\zeta_a=1-\frac{1}{z}$ 
	whose origin is at the horizon $\zeta_a=0$.  The differential equation has  only   one degenerate indicial root so that it has   a  logarithmic  solution.The regularity of the horizon  remove this log solution.  
	Putting $ y_{in}(\zeta_a) =\sum_{n=0}^{\infty} t_n \zeta_a^n$ into a differential equation, we obtain the four term recurrence relation as one can see in supplementary material,  and the convergence regime is 
	$|\zeta_a|<1$.   
	In Fig.~\ref{domain1}(b),  notice that the boundary point  of the  convergence regime is at $\zeta_a=-1$ which is    outside the black hole  and at the curvature singularity.  Now, if we impose the well-definedness at the boundary of the convergence,   $\zeta_a=-1$ or $r=2$, it   again  imply  the regularity at the black hole singularity which is located at $\zeta_a=1$.  Furthermore,   the locus in  $\Delta $-$\lambda$ plane   exactly overlap with that of the $\det M=0$ which we met in $a=1/2$ case.
	\paragraph*{  $\bullet \; a=0$: the  original coordinate:}\label{horowitz3}  
	
	Finally we come to  the coordinate centered at the curvature singularity, the original $r$ coordinate.   It is the boudanry of the   regime and the analysis is slightly    different. 
	In this case, the indicial root is degenerate, hence  the second solution contains  $\log z$. If we allow the presence of such second solution, even after we impose the finiteness at the  horizon, its derivative diverges there. To respect the equivalence principle, we have to request that the solution should be smooth at the horizon. 
	The horizon regularity as the smooth connection of the inside and outside solutions requests that the  second   solution should be removed to avoid the divergence of the first derivative. 
	After this imposition,  imposing  the horizon regularity to the rest of the solution,  is  to remove the  $\sum 1/n$, and  the locus in  $\Delta $-$\lambda$ plane   exactly overlap with that of the $\det M=0$ which appeared in the coordinate centered at the  regular point $r=1/3$.     
	For more detail, see the supplementary material. 
	The original work by Horowits et.al \cite{Hartnoll:2020fhc} used   $r$ coordinate  but did not considered smooth matching condition and therefore did not get finiteness at the singularity from the horizon regularity. 
	In the next paragraph, we consider $a\to 0$ limit instead of $a=0$ and show how things are different. 
	\paragraph*{  $\bullet \; a\rightarrow 0$:}\label{horowitz3} 
	We consider $a \rightarrow 0$ in eq.(\ref{gen}) and $A_n, B_n, C_n, D_n$  in  eq.(\ref{mm:200}) share $a(a^3-1)$ as their denominator. We can rewrite 
	\begin{footnotesize}
		\begin{equation}
			0=\det \mbox{M}_{N\times N}(\Delta, \lambda, a)=\frac{\sum_{n=0}^{2N} Z_n(\Delta, \lambda) a^n}{a^N (a^3-1)^N}
		\end{equation}
	\end{footnotesize}
	in the limit as $N \rightarrow \infty$. 
	We found that 
	the locus of $Z_i(\Delta, \lambda)=0$ at $i\in \{0,1,2,\cdots\}$ in     $\Delta $-$\lambda$ plane   exactly  overlap with that of the $\det M=0$ which appeared in the coordinate centered at the  regular point $r=1/3$.  This is independent of the value of $a\in (0,1)$. This  result is also true for in the case of   $a\to 1$. See the  fig.~\ref{intersection1}(b). \\

	\paragraph*{  Summary:}
	In all coordinate $\zeta_a=a-\frac{1}{z}$ with  $0 \leq a \leq 1$, the horizon regularity condition with well-definedness implies the regularity at the center of a black hole.  As a result, $\Delta$ and $\lambda$ are quantized and  independent of $a$.   Our method can be easily generalized to   operators other than the superconducting order parameter. 	Several examples are planned to appear later. \\
	
	
	
	\paragraph*{Acknowledgments: } 
	The  work of YSC, TWY and SJS is supported by Mid-career Researcher Program through the National Research
	Foundation of Korea grant No. NRF-2021R1A2B5B02002603, RS-2023-00218998 and NRF-2022H1D3A3A01077468.  The work of KSK is supported by the National Research Foundation of Korea (NRF) grant   NRF- 2021R1A2C1006453 and  NRF-2021R1A4A3029839. 
	We thank the APCTP for the hospitality during the focus program, where part of this work	was discussed.
	
	%
	%
	\bibliography{Refs_SC}


  \newpage 	
	
	 	\onecolumngrid	 

		\section*{Supplementary Material}
	
	\section{Preliminaries:} 
		Before we explain the scenario for solving the problem, we will first present the well known theorem.
		\begin{thm}
			Birkhoﬀ–Adams theorem \cite{elaydi1996introduction}:
			
			Consider  the  second-order  diﬀerence  equation
			\begin{equation}
				d_{n+1}+ p_n \;d_n + q_n \;d_{n-1}=0,   \quad
				\label{eee:5}
			\end{equation}
			where $ p_n$  and $ q_n$  have  asymptotic  expansions  of  the  form
			\begin{equation}
				\begin{cases}  p_n\sim \sum_{j=0}^{\infty}\frac{a_j}{n^j} , \cr
					q_n\sim \sum_{j=0}^{\infty}\frac{b_j}{n^j} . 
				\end{cases}
				\label{eee:6}
			\end{equation}
			The  characteristic  equation  associated  with  eq.(\ref{eee:5})  is $ \rho^2  + a_0\; \rho + b_0 =0$ with  roots
			\begin{equation}
				\rho_1, \rho_2= -\frac{a_0}{2} \pm \sqrt{\frac{a_0^2}{4}-b_0}.  \quad
				\label{eee:7}
			\end{equation}
			If $\rho_1 \ne \rho_2$ ,  i.e.,$a_0^2 \ne 4b_0$, then eq.(\ref{eee:5}) has two linearly independent solutions $ d_{n}^{(1)},  d_{n}^{(2)}$,   which   will   be   called   normal   solutions.  
			They have   the  form  
			\begin{equation}
				d_{n}^{(i)} \sim \rho_i^{n} n^{\alpha_i}\sum_{r=0}^{\infty}\frac{c_i(r)}{n^r} \quad
				\label{eee:8}
			\end{equation}
			\begin{equation}
				\alpha_i= \frac{a_1 \rho_i +b_1}{a_0 \rho_i +2 b_0} \quad
				\label{eee:9}
			\end{equation}
			with $c_i(0)=1$.
			In particular, we have
			\begin{small}
				\begin{equation}
					c_i(1)=\frac{-2\rho_i^2\alpha_i(\alpha_i-1)-\rho_i (a_2+a_1 \rho_i +\alpha_i(\alpha_i-1)a_0/2 -b_2)}{2\rho_i^2(\alpha_i-1)+\rho_i (a_1+(\rho_i-1)a_0)+b_1}\quad
					\nonumber
				\end{equation}
				We can apply this theorem into the higher-order linear difference equation.
			\end{small} 
		\end{thm}

		\section{ Near critical temperature  for 2+1 dimensional system:
		}\label{gaga} 
		We  take the ansatz  that the scalar field depends only on the holographic direction to have  the form $ \Psi (z)$ and  electrostatic   potential is also of the form $ \Phi(z)$. 
		The Maxwell scalar field equations with     AdS$_{d+1}$-Schwarzschild blackhole  background become  
		\begin{small}
			\bea
			&&  \frac{d^2 \Psi }{d z^2} -\frac{d-1+z^d}{z(1-z^d)}\frac{d \Psi }{d z}+\left[ \frac{g^2 \Phi^2}{r_h^2(1-z^d)^2}-\frac{m^2}{z^2(1-z^d)}\right]\Psi =0, \nonumber \\
			&& \frac{d^2 \Phi}{d z^2}-  \frac{d-3}{z}\frac{d \Phi}{d z } -\frac{2g^2\Psi^2}{z^2(1-z^d)}\Phi =0 
			\label{qqq:1}
			\eea
		\end{small}
		with the coordinate $z= r_h/r$  with setting $r_h=1$.
		
		In terms of methodology, we will divide the domain of the solution   into two regions, one is the outside ($|z|\leq 1$)  and the other is the inside  ($|z|\geq 1$) of the black hole, and then smoothly  connect the solutions over these two regions  at the event horizon at $z=1$: Generically two solutions do not converge at $z=1$; however, if we can kill the dominant solution which diverges at the horizon then 
		two solutions can be joined smoothly there. 
		
		At the critical temperature $T_c$, the condensation can be set  to be zero, $\Psi =0$, so   Eq.(\ref{qqq:1}) at $d=3$ tells us $\Phi^{''}=0$. Then, we can set
		\begin{equation}
			\Phi(z)= \tilde{\lambda}  r_c (1-z) \hspace{1cm}\mbox{where}\;\;\tilde{\lambda} =\frac{\rho}{r_c^2}
			\label{mm:9}
		\end{equation}
		As $T\rightarrow T_c$, the field equation $\Psi$ approaches to   
		\begin{equation}
			-\frac{d^2 \Psi }{d z^2} +\frac{2+z^3}{z(1-z^3)}\frac{d \Psi }{d z}+  \frac{m^2}{z^2(1-z^3)} \Psi =\frac{\lambda^2}{(z^2+z+1)^2}\Psi
			\label{eq:10}
		\end{equation}
		where $\lambda = g\tilde{\lambda}$.
		Factoring out the behavior near the boundary $z=0$ and those near  $z=\frac{ -1\pm i \sqrt{3} }{2} $, we define 
		\begin{eqnarray}
			&&\Psi(z)= \frac{\left< \mathcal{O}_{\Delta}\right>}{\sqrt{2}r_h^{\Delta}}z^{\Delta} F(z) \nonumber \\&&\mbox{where}\;\;F(z)=
			(z^2+z+1)^{-\lambda/\sqrt{3}} Y_{out}(z).
			\label{mm:1}
		\end{eqnarray}    
		Then, $F$ is normalized as $F(0)=1$ and we obtain
		\begin{equation}
			\frac{d^2 Y_{out} }{d z^2} +\frac{a_3 z^3+a_2 z^2 +a_1 z +a_0}{z( z^3-1)}\frac{d Y_{out}}{d z}   +\frac{b_2 z^2 + b_1 z +b_0}{ z(
				z^3-1)}Y_{out} =0, 
			\label{mm:2}
		\end{equation}
		\begin{equation}
			\hbox{where }		\begin{cases} 
				a_0= 2(1-\Delta) , \quad 
				a_1=\frac{2\lambda }{\sqrt{3}} , \quad 
				a_2=\frac{2\lambda }{\sqrt{3}}  ,\cr 
				a_3=  1-\frac{4\lambda}{\sqrt{3}}    +2\Delta ,\cr 
				b_0=-\frac{2\lambda}{\sqrt{3}} (1-\Delta) , \cr
				b_1= \frac{1}{3}(-4\lambda^2+2\sqrt{3}\Delta\lambda -\sqrt{3}\lambda ), \cr
				b_2= \frac{1}{3}(3\Delta^2-4\sqrt{3}\Delta\lambda +4\lambda^2 ).        
			\end{cases}
			\label{mmmm:2}
		\end{equation} 
		This   equation is a generalized Heun's equation that has five regular singular points at $z=0,1,\frac{-1\pm\sqrt{3}i}{2},\infty$.
		Substituting $ Y_{out}(z)= \sum_{n=0}^{\infty } d_n z^{n}$ into eq. (\ref{mm:2}), we obtain the four term  recurrence relation:
		\begin{equation}
			d_{n+1}= A_n \;d_n + B_n \;d_{n-1}+C_n\;d_{n-2}  \label{mm:3}
		\end{equation}
		$\hbox{  for  } n \geq 2$. Here, $A_n, B_n, C_n $ consists with $\Delta$,  $\lambda$ and index $n$.
		with  
		\begin{equation}
			\begin{cases}  A_n=\frac{2\sqrt{3}\lambda(n+\Delta-1)}{3(n+1)(n+2\Delta-2)}\cr 
				B_n=\frac{\sqrt{3}(2n+2\Delta-3)\lambda-4\lambda^2}{3(n+1)(n+2\Delta-2)} \quad 		C_n=\frac{(n-\frac{2}{\sqrt{3}}\lambda+\Delta-2)^2}{(n+1)(n+2\Delta-2)}, 
			\end{cases}
			\label{mm:4}
		\end{equation} 
		and the first three  $d_{n}$'s are  $d_{0}=1$, $d_1=A_{0}d_0$ and  
		$d_{2}=A_{1}d_1 +B_{1}d_{0}$. 
		Eq.(\ref{mm:1}), Eq.(\ref{mm:3}) and Eq.(\ref{mm:4})  give us the   boundary condition
		$		F^{'}(0)=0.
		\nonumber
		$	
		
		
		Now,  $d_n$ has three   independent solutions $	d_{n}^{(1)}$,  $	d_{n}^{(2)}$ ,  $	d_{n}^{(3)}$.
		Applying the  Birkhoﬀ–Adams theorem
		\cite{elaydi1996introduction},  such that 
		$
		d_n  \sim  a_0	d_{n}^{(1)} + a_1	d_{n}^{(2)}+ a_2	d_{n}^{(3)} $ for sufficiently large $n$ where
		\begin{equation}
			\begin{cases}   d_{n}^{(1)}\sim n^{-1}  \cr
				d_{n}^{(2)}\sim  \left( \frac{-1+\sqrt{3}i}{2} \right)^n  n^{-1-\frac{2\lambda }{\sqrt{3}}}   \cr
				d_{n}^{(3)}\sim  \left( \frac{-1-\sqrt{3}i}{2} \right)^n n^{-1-\frac{2\lambda }{\sqrt{3}}} . 
			\end{cases}
			\nonumber
		\end{equation} 
		Here, $a_0, a_1, a_2$ consist of $\Delta$ and $\lambda$.  Here, $	d_{n}^{(2)}$ is  the complex conjugate of $	d_{n}^{(3)}$. $d_n$ should be real because all coefficients are real in it. So, $a_1=a_2$ must to be valid.
		Therefore, 
		$
		d_n \sim  a_0	d_{n}^{(1)} + a_1	d_{n}^{(2)} $  for sufficiently large $n$ where
		\be
		d_{n}^{(1)}=  n^{-1} ,  \quad 
		d_{n}^{(2)}= \frac{    \cos \left(\frac{2 \pi  n}{3}\right)}{n^{\frac{2 \lambda }{\sqrt{3}}+1}}.\nonumber\ee

		This means that a harmonic series   $\sum_{n}^{\infty}d_n^{(1)} z^n$ develops $a_{0} \log(1-z)$ behavior near horizon. 	 
		For the horizon regularity, we require $a_{0}=0$. 
		With this condition, it makes convergent at the horizon.  
	$a_0=0$ is equivalent to 
	\begin{equation}
		\lim_{n\to \infty}  (n+1) d_{n+1}=0  
		\label{mm:12}
	\end{equation}  
	The solution of such equation is given by the  Fig.~\ref{intersection1}.  In Fig.~\ref{intersection1} (b),  red curves are solutions of the transcendental equation eq.(\ref{mm:12})  of $\lambda$ with given $\Delta$ for outside of balck hole.  

\vskip .3in 
Now we consider inside of a black hole. We change a coordinate $\xi= 1/z$ in  eq.(\ref{mm:2}).    
Factoring out the behavior near $\xi =0$, the general solution is taken as

\begin{equation}
Y_{in}(\xi) = \mathcal{A}  \xi^{\Delta -\frac{2 \lambda }{\sqrt{3}}} \left(  A G_{1}(\xi) +B G_{2}(\xi) \right)
\label{ttt:1}
\end{equation}  
Here,
\begin{equation}
G_{1}(\xi)=   \sum_{n=0}^{\infty } d_n \xi^{n},
\label{dis:2}
\end{equation}
is called as the first kind of a series.
And,
\begin{eqnarray}
G_{2 }(\xi) &=& G_{1}(\xi) \log(\xi) + \widetilde{G}(\xi) \nonumber 
\label{dis:3}
\end{eqnarray}
called as the second kind of a soultion.
Notice that the series has only one   indicial root, thus  $\log(\xi)$ term is arisen.

From Birkhoﬀ–Adams theorem, we have the following asymptotic expression; 
\bea
&&  \lim_{\xi \to  1} G_{1}(\xi) = a_0  \log(1-\xi) ,  \label{tt:5}\\
&& \lim_{\xi \to  1} \widetilde{G}(\xi)  = a_1  \log(1-\xi)\label{tt:6}.
\eea	
Here, $a_0, a_1$ consist of ($\Delta$, $\lambda$). And
\bea  &&  \lim_{\xi \to 1} \Psi(\xi)  =\lim_{\xi \to 1}  \left( A G_{1}(\xi) +B G_{2}(\xi) \right)   = \lim_{\xi \to 1}  \left( A G_{1}(\xi) +B   \widetilde{G}(\xi) \right) \nonumber \\
&& \sim  \left( A a_0  +B  a_1 \right) \log(1-\xi)\eea 
According to the horizon regulation condition, we must set $B= -\frac{A a_0 }{a_1 }$, $\log(1-\xi)$ behavior does not exist any more. This process makes that  $A G_{1}(\xi) +B   \widetilde{G}(\xi) $ converge at the horizon. However, if we take the derivative of $ \Psi(\xi) $, it is not convergent any more. Because  
\begin{eqnarray} 
&& \lim_{\xi \to 1}\frac{d  \Psi(\xi) }{d \xi }= \lim_{\xi \to 1}\frac{d  }{d \xi } \Big(  A G_{1}(\xi) +B G_{2}(\xi) \Big)  \nonumber\\
&=& \lim_{\xi \to 1} \frac{d  }{d \xi } \Big(  B G_{1}(\xi) \log \xi    \Big)+ \lim_{\xi \to 1} \frac{d  }{d \xi } \Big( A G_{1}(\xi) +B   \widetilde{G}(\xi)  \Big) \nonumber\\
&&\sim \lim_{\xi \to 1} \frac{d  }{d \xi } \Big( B G_{1}(\xi) \log \xi  \Big) \rightarrow \infty.
\end{eqnarray} 
Notice that  we make $A G_{1}(\xi) +B   \widetilde{G}(\xi)  $ convergent at $\xi=1$ on the above by setting $B= -\frac{A a_0 }{a_1 }$. In other words, it is analytic at the horizon and  differentiable at that point. So $$\lim_{\xi \to 1} \frac{d  }{d \xi } \Big( A G_{1}(\xi) +B   \widetilde{G}(\xi)  \Big) < \infty.$$
Therefore, we should remove $\log \xi$ term at $\xi=\infty$ by setting $B=0$. To make horizon regularity, $a_0(\Delta, \lambda) =0$ should be also satisfied. Here, $a_0 $ consists of $\Delta$ and $\lambda$, and this constrained condition give us the quantization of $\Delta$.	

The new  eq.(\ref{ttt:1}) is given by
\begin{equation}
Y_{in}(\xi)= \mathcal{A}\xi^{\Delta -\frac{2 \lambda }{\sqrt{3}}} G_{1}(\xi). 
\label{mm:14}
\end{equation}
here, $\mathcal{A}$ is a connection coefficient which can be determined by 
requesting the smooth connection conditions:  
\be
Y_{out}(1)=Y_{in}(1), \quad Y_{out}'(1)=Y_{in}'(1).\label{bc111}\ee 
Finally    we obtain  
\begin{equation}
\frac{d^2 G_{1} }{d \xi^2} +\frac{g_3 \xi^3+g_2 \xi^2 +g_1 \xi +g_0}{\xi( \xi^3-1)}\frac{d G_{1}}{d \xi}   +\frac{h_2 \xi^2 + h_1 \xi +h_0}{ \xi(
	\xi^3-1)}G_{1} =0, 
\label{mm:15}
\end{equation}
\begin{equation}
\hbox{where }		\begin{cases} 
	g_0=-1, \quad 
	g_1=\frac{2 \lambda }{\sqrt{3}} , \quad 
	g_2= \frac{2 \lambda }{\sqrt{3}} ,\cr 
	g_3= 4 \left(1-\frac{\lambda }{\sqrt{3}}\right)  ,\cr 
	h_0= \frac{\lambda }{\sqrt{3}} , \cr
	h_1= \frac{2}{3} \lambda  \left(\sqrt{3}-2 \lambda \right) , \cr
	h_2= -\Delta  (\Delta -3)+\frac{4 \lambda ^2}{3}-2 \sqrt{3} \lambda.        
\end{cases}
\nonumber
\end{equation} 
This is also a generalized Heun's equation  that has five regular singular points at $\xi=0,1,\frac{-1\pm\sqrt{3}i}{2},\infty$.\\
Substituting $ G_{1}(\xi)= \sum_{n=0}^{\infty } s_n \xi^{n}$ into eq. (\ref{mm:15}), we obtain the four term  recurrence relation:
\begin{equation}
s_{n+1}= A_n \;s_n + B_n \;s_{n-1}+C_n\;s_{n-2}   \quad
\hbox{  for  } n \geq 2, \label{mm:16}
\end{equation}
with
\begin{equation}
\begin{cases}  
	A_n=  \frac{\lambda  (2 n+1)}{\sqrt{3}(n+1)^2} \cr
	B_n= \frac{2 \lambda  \left(n-\frac{2 \lambda }{\sqrt{3}}\right)}{\sqrt{3}(n+1)^2}  \cr
	C_n=  \frac{-(\Delta -3) \Delta +\frac{4 \lambda ^2}{3}-2 \sqrt{3} \lambda +(n-2) \left(-\frac{4 \lambda }{\sqrt{3}}+n+1\right)}{(n+1)^2}
\end{cases}
\label{mm:17}
\end{equation}   
And the first three  $s_{n}$'s are  $s_{0}=1$, $s_1=A_{0}s_0$ and  
$s_{2}=A_{1}s_1 +B_{1}s_{0}$.  
Substitute eq. (\ref{mm:17}) in eq. (\ref{mm:12}). The new eq. (\ref{mm:12}) should be applied for this case.

For large $n$, $s_{n} \sim a \; s_{n}^{(1)} +b\; s_{n}^{(2)}+c\; s_{n}^{(3)}$. Here, $a,b,c$ are some constants, and $ s_{n}^{(i)}$ is an independent solution of $s_n$. 

Similarly to the previous case, we obtain  
\begin{equation}
\begin{cases}   s_{n}^{(1)}\sim n^{-1}  \cr
	s_{n}^{(2)}\sim  \left( \frac{-1+\sqrt{3}i}{2} \right)^n  n^{-1-\frac{2\lambda }{\sqrt{3}}}   \cr
	s_{n}^{(3)}\sim  \left( \frac{-1-\sqrt{3}i}{2} \right)^n n^{-1-\frac{2\lambda }{\sqrt{3}}} . 
\end{cases}
\label{mm:18}
\end{equation}
Here, $	s_{n}^{(2)}$ is  the complex conjugate of $	s_{n}^{(3)}$. $s_n$ should be real because all coefficients are real in it. So, $b=c$ must to be valid.
Therefore, 
$
s_n   \sim  a s_{n}^{(1)} + b 	s_{n}^{(2)} $ for large $n$  where
\be
s_{n}^{(1)}=  n^{-1} ,  \quad 
s_{n}^{(2)}= \frac{    \cos \left(\frac{2 \pi  n}{3}\right)}{n^{\frac{2 \lambda }{\sqrt{3}}+1}}.\nonumber\ee 
This means that a harmonic series   $\sum_{n}^{\infty}s_n^{(1)} \xi^n$ develops $a  \log(1-\xi)$ behavior near horizon. 	 
For the horizon regularity, we require $a=0$. With this condition,  it makes convergent at the horizon. Similar to the previous case for outside solution, $	\lim_{n\to \infty}    (n+1) s_{n+1}  =0 $  should be required for the regularity at the horizon. 
\begin{figure}[!htb]
\subfigure[]
{ \includegraphics[width=0.4\linewidth]{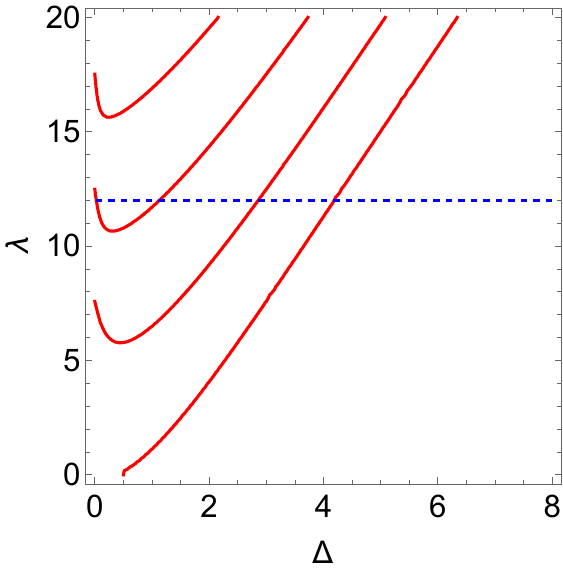}}
\subfigure[]
{ \includegraphics[width=0.4\linewidth]{Ads4_v1.pdf}}  
\caption{   $\lambda $  vs $\Delta$. 
	(a) for the fixed charge density and coupling,  the scaling dimension  can have only quantized value even when we look at the outside solution only. (b)
	Red and Blue curves are solutions   for outside and inside of balck hole respectively. 	If we request that they are smoothly connected at the horizon only the intersection points are allowed parameters. 	} 
\label{intersection1}
\end{figure}  
Blue curves show the solution of such equation for inside of the black hole in the  Fig.~\ref{intersection1} (b).   
We emphasize that   for a given $\lambda$, which is the density times coupling, we can find discrete set of scaling dimension $\Delta$ as one can see in figure \ref{intersection1} (a).    That is, the scaling dimension  can have only quantized value even when we look at the outside solution only if we fix $\lambda$.  


All intersection points represent the solution of the new eq.(\ref{mm:12}).  
To list the first a few of them,  
\\$ (\Delta,\lambda)=(3/2,2.477), (3/2,7.723), (3/2,12.939),  \cdots, $ 
which are all lying on the line $\Delta=3/2$. 
For   $\Delta< 3$    there are only two  more intersection points    $ (\Delta,\lambda)=(2.881, 7.156),\; (2.997,12.477)$ other than the $\Delta=3/2$ series.

One  interesting consequence is that our solution of scalar field given in eq.(\ref{mm:1})  always has asymptotic behavior which saturate to the finite  constant although we never requested its finiteness:
\begin{equation}
\Psi(z\to \infty)  
\simeq  \frac{\left< \mathcal{O}_{\Delta}\right>}{\sqrt{2}r_h^{\Delta}}  \mathcal{A}=\frac{\left< \mathcal{O}_{\Delta}\right>}{\sqrt{2}r_h^{\Delta}}  \frac{F(1)}{ G_{1}(1)} .
\label{qq:111}
\end{equation}

\begin{figure}[!htb] 
\subfigure[]
{ \includegraphics[width=0.4\linewidth]{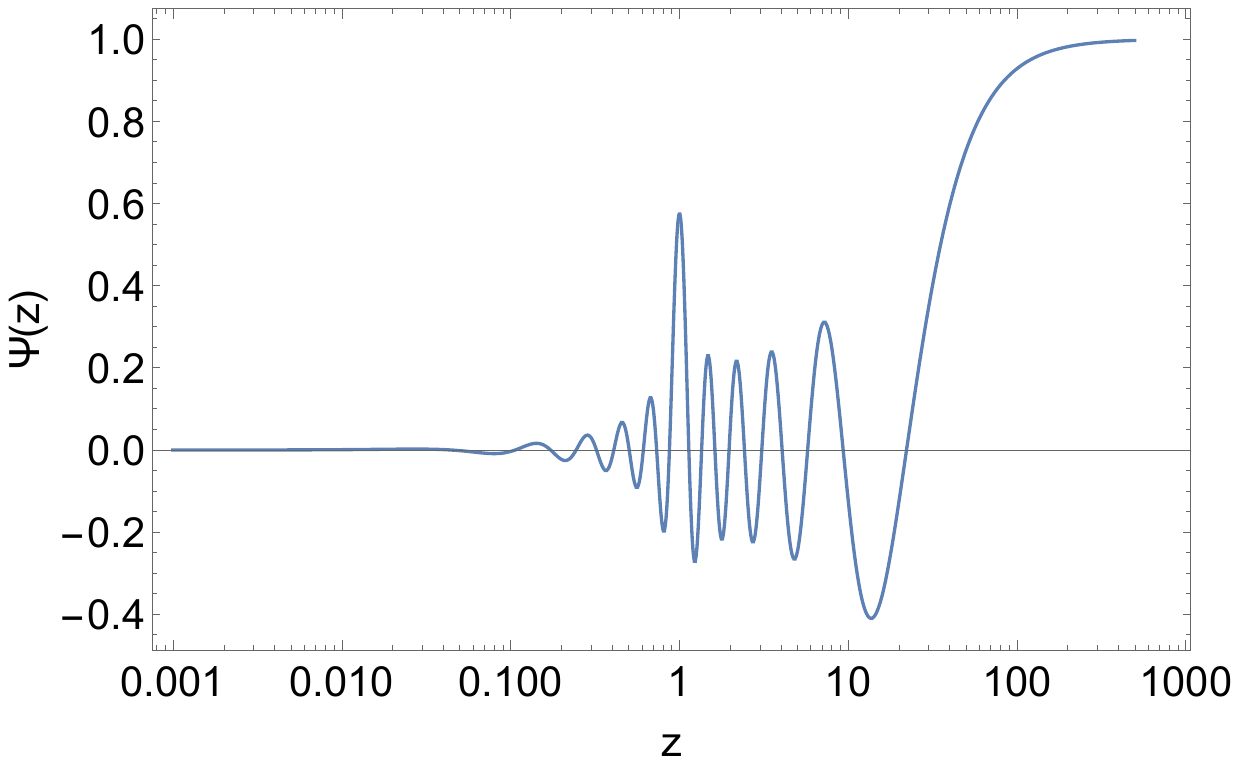}}  
\subfigure[]
{ \includegraphics[width=0.4  \linewidth]{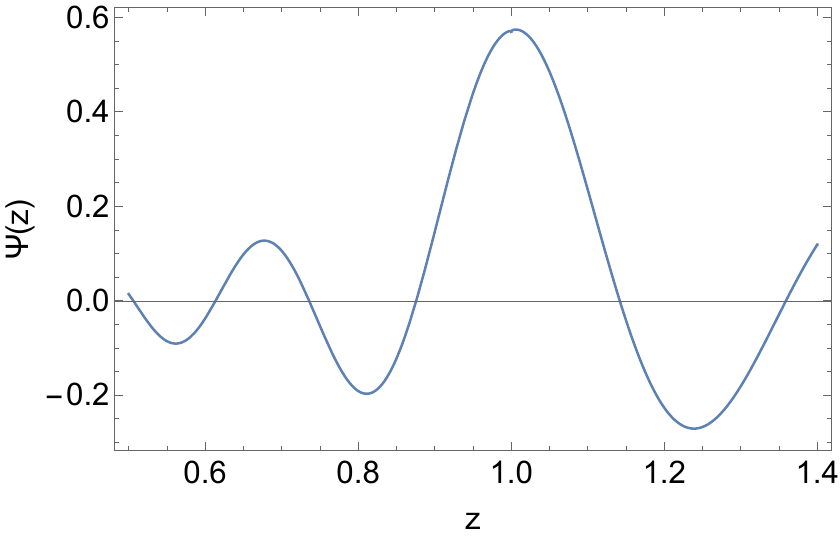}}  
\caption{ $\Psi(z)$ for  different values of $(\Delta, \lambda)$. 
	We plot  numerically with given $(\Delta, \lambda)$, obtained by  Fig.~\ref{intersection1}. Here, we set $\frac{\left< \mathcal{O}_{\Delta}\right>}{\sqrt{2}r_h^{\Delta}}=1$ for convenience.
	(a) As $\lambda$ is bigger with given $\Delta$, the solution oscillate rapidly. Here, we have $(\Delta, \lambda)=(3/2, 54.55)$.  (b) Zoom  in of figure (a) near $z=1$, showing the smooth connection at the horizon, $z=1$.} 
\label{polynomial}
\end{figure}

One  interesting consequence is that our solution of scalar field given in eq.(\ref{mm:14})  always has asymptotic behavior which saturate to the finite  constant $\Psi(z\to \infty) < \infty$. 
We plot  solutions $\Psi(z)$ for different values of $(\Delta, \lambda)$'s in Fig.~\ref{polynomial}. Recently,  Hartnoll et.al \cite{Hartnoll:2020fhc} studied  the  solution  of holographic superconductor inside black hole   from the different perspective; their solution of scalar field near critical temperature has the oscillation of $\Psi(z)$ inside of black hole. Our result is  the solution to the linearized equation. Nevertheless, we have a  similar oscillation for large $\lambda$ with given $\Delta$ in  (a) in Fig.~\ref{polynomial}. The   differences are i) $\Psi(z)$ is finite at the center of black hole, ii) $\Delta$ and $\lambda$ values are quantized. 

Summarzing, imposing the equivalnece principle at the horizon of the black hole lead to the quanzation of the allowed value of scaling dimension.

As we see  (b) in Fig.~\ref{polynomial}, numerical results tell us that  two plots of outside and inside solutions are connected to each other at the horizon smoothly.

\section{Physical aspects in the different coordinate system:} 

we   introduce  one parameter family  of  coordinates 
\be 
\zeta_a=a-\frac{1}{z}, 
\ee
centered at $r=a$  with    $0\leq a \leq1$.
We change the coordinate $\zeta_a$  in eq.(\ref{mm:2}). And take $Y_{in}(\zeta_a) = \left( a-\zeta_a\right)^{\Delta -\frac{2 \lambda }{\sqrt{3}}}y_{in}(\zeta_a)$ into the new  eq.(\ref{mm:2}).    

Expanding   $ y_{in}(\zeta_a)= \sum_{n=0}^{\infty } e_n \zeta_a^{n}$ and insert it to the new differential equation,   we obtain the five term  recurrence relation.
we have
\begin{equation} 
e_{n+2}=	A_n e_{n+1}+B_n e_n+C_n e_{n-1}+D_n e_{n-2}  \label{gen:1}
\end{equation}  
$\hbox{  for  } n \geq 2$, 
where
\begin{equation}
\begin{cases}  
A_n=  \frac{12 a^3 (n+1)+2 \sqrt{3} \left(-2 a^2+a+1\right) a \lambda -3 (n+1)}{3 a \left(a^3-1\right) (n+2)} \cr
B_n=  -\frac{-3 a^2 \Delta ^2+9 a^2 \Delta +18 a^2 n (n+1)+4 (a-1) a \lambda ^2+2 \sqrt{3} \lambda  (n-a (3 a-1) (2 n+1))+\sqrt{3} \lambda }{3 a \left(a^3-1\right) (n+1) (n+2)}  \cr
C_n= \frac{2 \left(a \left(-3 (\Delta -3) \Delta +4 \lambda ^2+6 n^2-6 \sqrt{3} \lambda  n-6\right)+\lambda  \left(\sqrt{3} n-2 \lambda \right)\right)}{3 a \left(a^3-1\right) (n+1) (n+2)}  \cr
D_n = \frac{3 (\Delta -3) \Delta -2 \lambda  \left(2 \lambda +\sqrt{3}\right)-3 n^2+\left(4 \sqrt{3} \lambda +3\right) n+6}{3 a \left(a^3-1\right) (n+1) (n+2)}
\end{cases}
\nonumber
\end{equation}  
The first two  $e_{n}$'s are  $e_2=   A_0 e_1+B_0 e_0$ and  
$e_3= A_1 e_2+B_1 e_1+C_1 e_0$.
\eqref{gen:1} can be represented in the matrix form as
\begin{equation}
\sum_{m=0}^{\infty} \mbox{M}_{n,m} \; e_m =0 \hspace{1cm}  (n=0, 1, 2, \cdots), \label{gen:2}
\end{equation} 
where  	$\mbox{M}_{n,n}=B_n $, 
$	\mbox{M}_{n,n+1}=A_n$, 
$\mbox{M}_{n,n+2}=-1$, 
$\mbox{M}_{n+1,n}=C_{n+1}$ and $	\mbox{M}_{n+2,n}=D_{n+2} $.
To handle this infinite dimensional matrix approximately, we first truncate to $n=N+2$ for large enough $N$. 
If we set $e_{N+1}=0$ and $e_{N+2}=0$,  $\mbox{M} $ becomes $N \times N$ matrix.	Only when $\det\mbox{M}  = 0$ is satisfied, i.e.,
\begin{equation}
\det\mbox{M}=\begin{vmatrix}
B_0& A_0& -1 &  &  &  &  &     \\
C_1 & B_1  &  A_1 &  -1  &  &  &  &    \\
D_2  & C_2  & B_2  & A_2  &  -1 &  &  &     \\
& D_3 & C_3   & B_3 &   A_3 &  -1 &  &    \\
&  & D_4 &  C_4 & B_4 &  A_4 &  -1 &     \\
&  &  & \ddots & \ddots & \ddots & \ddots &  \ddots    \\
&  &  &  &D_{n-1}  &C_{n-1}  &B_{n-1}  &A_{n-1}   \\
&  &  &  &  & D_{n} & C_{n} & B_{n} 
\end{vmatrix}=0, 
\label{gen:3}
\end{equation}   
\eqref{gen:3} allows a nontrivial solution. The converse is also true, because, with truncation, if $\det\mbox{M}_{N \times N}=0$, then  $e_{N+1}=0$, $e_{N+2}=0$. Therefore,
\begin{equation}
e_{N+1}=0, e_{N+2}=0 \leftrightarrow  \det\mbox{M}_{N \times N}=0.
\label{main}
\end{equation}

Now,  for large $n$, $e_n$  can be written as sum of three terms  $	e_{n}^{(1)}$,  $	e_{n}^{(2)}$,  $e_{n}^{(3)}$
\cite{elaydi1996introduction}: 
$  e_n  \sim  \nu_0	e_{n}^{(1)} + \nu_1	e_{n}^{(2)}+ \nu_2	e_{n}^{(3)} $ where
$$ e_{n}^{(1)}=  \frac{\left(\frac{1}{a-1}\right)^n}{n},  e_{n}^{(2)}= \frac{\left(\frac{1}{a}\right)^n}{n}, e_{n}^{(3)}=\frac{\left(\frac{1}{a^2+a+1}\right)^{n/2} \cos (\alpha  n)}{n^{\frac{2 \lambda }{\sqrt{3}}+1}} $$ with 
$\alpha =\tan ^{-1}\left(\frac{\sqrt{3}}{2 a+1}\right)$. Here, $\nu_0, \nu_1, \nu_2$ consist of $\Delta$, $\lambda$ and $a$.
On the other hand, 
$e_n$ can be  written as a linear combination of $e_0$ and $e_1$: 
\be 
e_n  = e_0	V_n(\Delta,\lambda)   + e_1 W_n(\Delta,\lambda),
\label{gen:4}
\ee
because   all the rest of $e_n$ are given as   polynomials of  $A_n,B_n,...$ and  it is linear in $e_0,e_1$.  Then $	 e_{N+1}=0, \;   e_{N+2}=0 $  is equivalent to   
\be 
V_{N+2} W_{N+1}-V_{N+1} W_{N+2}=0. 
\label{gen:5}
\ee
Since   $\nu_2	e_{n}^{(3)} $  is negligible for    sufficiently large $n$,   we can say
\be    \lim_{n\to \infty}e_n = \nu_0	e_{n}^{(1)} + \nu_1	e_{n}^{(2)}.
\ee 
Then $\nu_0$ and $\nu_1$ are  given  as
\begin{eqnarray}
\nu _0 &=&   \frac{(-1)^n \left((n+2) e_{n+2}-2 (n+1) e_{n+1}\right)}{2^{n+3}} \nonumber\\
\nu _1 &=& \frac{2 (n+1) e_{n+1}+(n+2) e_{n+2}}{2^{n+3}}
\label{gen:6} 
\end{eqnarray}
in the limit  $n \to \infty$. Substitute \eqref{gen:4} into  \eqref{gen:6}, we see 
\be 
\nu_0=\nu_1=0  \leftrightarrow V_{n+2} W_{n+1}-V_{n+1} W_{n+2}=0. 
\label{gen:7}
\ee  
From  \eqref{gen:5},\eqref{gen:7} ,  $\nu_0= \nu_1=0$ is equivalent to 
$e_{N+1}= e_{N+2}=0 $. Then by \eqref{main},
\be 
\nu_0=0, \;  \nu_1=0  \leftrightarrow  \det \mbox{M}  = 0.
\label{main1}
\ee  
\subsection{ The simplest working case:} 
We change the coordinate $\zeta_a $ at $a=1/2$ in eq.(\ref{mm:2}).
\begin{figure}[!htb]
\centering
{ \includegraphics[width=0.3\linewidth]{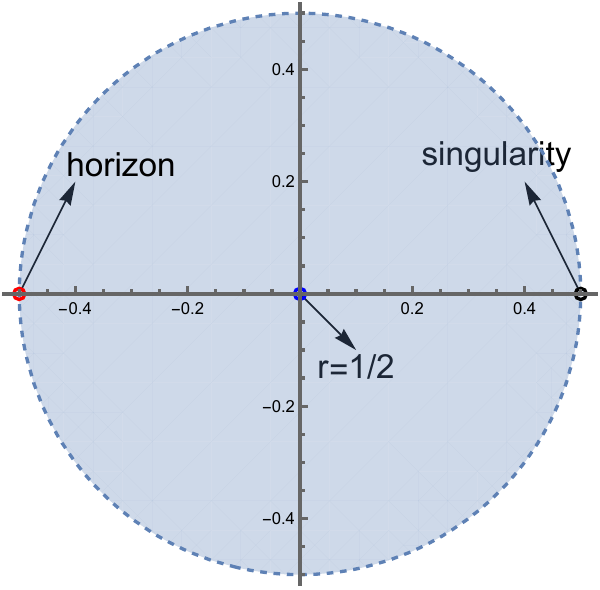}} 
\caption{ Domain of $y_{in}(\zeta_{a})$ at $a=1/2$.   } 
\label{domain}
\end{figure}  
And take $Y_{in}(\zeta_a) = \left( \frac{1}{2}-\zeta_a\right)^{\Delta -\frac{2 \lambda }{\sqrt{3}}}y_{in}(\zeta_a)$ into the new  eq.(\ref{mm:2}).    
Putting $a=1/2$ into \eqref{gen:1},
we have
\begin{equation} 
e_{n+2}=	A_n e_{n+1}+B_n e_n+C_n e_{n-1}+D_n e_{n-2}  \label{mm:200}
\end{equation} 
$\hbox{  for  } n \geq 2$, 
where
\begin{equation}
\begin{cases}  
A_n=  \frac{8 \left(-2 \sqrt{3} \lambda +3 n+3\right)}{21 (n+2)} \cr
B_n= \frac{4 \left(-3 (\Delta -3) \Delta -4 \lambda ^2+2 \sqrt{3} \lambda +4 \sqrt{3} \lambda  n+18 n (n+1)\right)}{21 (n+1) (n+2)}  \cr
C_n= \frac{16 \left(3 (\Delta -2) (\Delta -1)-6 n^2+4 \sqrt{3} \lambda  n\right)}{21 (n+1) (n+2)}  \cr
D_n =  \frac{16 \left(-3 (\Delta -3) \Delta +4 \lambda ^2+2 \sqrt{3} \lambda -4 \sqrt{3} \lambda  n+3 (n-1) n-6\right)}{21 (n+1) (n+2)}
\end{cases}
\nonumber
\end{equation}  
The first two  $e_{n}$'s are  $e_2=   A_0 e_1+B_0 e_0$ and  
$e_3= A_1 e_2+B_1 e_1+C_1 e_0$.

Now,  $e_n$ has three   independent solutions $	e_{n}^{(1)}$,  $	e_{n}^{(2)}$,  $e_{n}^{(3)}$
\cite{elaydi1996introduction},  such that 
$
\lim_{n\to \infty} e_n   =  \nu_0	e_{n}^{(1)} + \nu_1	e_{n}^{(2)}+ \nu_2	e_{n}^{(3)} $ where 
\be
e_{n}^{(1)}=  \frac{  (-2)^n}{n},  \quad 
e_{n}^{(2)}= \frac{  2^n}{n}, \quad e_{n}^{(3)}=  \frac{ \cos \left(m n  \right)\left(\frac{2}{\sqrt{7}}\right)^n}{n^{1+\frac{2 \lambda }{\sqrt{3}} }} \label{mm:202} \ee 
where $m=\arctan \left(\frac{\sqrt{3}}{2}\right)$.  

The dominant term in $\lim_{n\to \infty} e_n $ is  $  \frac{  (\pm 2)^n}{n}.$ 
From the ratio test, we obtain $|\zeta|<1/2.$ 
At the horizon,  \eqref{mm:202} gives us the asymptotic series for $y_{in}(\zeta)$ such as $\sum_{n }^{\infty} e_n \zeta^n  \sim   \nu_0 \sum _{n }^{\infty } \frac{1}{n}+  \nu_1 \sum _{n }^{\infty } \frac{\left(-1 \right)^n}{n}$. Similarly, at the singularity,  $\sum_{n }^{\infty} e_n \zeta^n  \sim  \nu_0 \sum _{n }^{\infty } \frac{\left(-1 \right)^n}{n}+  \nu_1 \sum _{n }^{\infty } \frac{1}{n}$. 
For the regularity at the horizon, we require $\nu_0 =0$. Here, a harmonic series $\sum _{n }^{\infty } \frac{1}{n}$ diverges in which has $\log$ divergence.   
The sum  $\sum _{n }^{\infty } \frac{\left(-1 \right)^n}{n}$ is called as alternating harmonic series that is  conditionally convergent. 	According to Riemann,   such conditionally convergent 
series may be rearranged to converge to any value  from $-\infty$ to $-\infty$.
Therefore, we accept the well-definedness of the series solution with setting $\nu_1 =0$. Then, the regularization at the singularity automatically satisfies. This means that the horizon regularity implies the regularity at the center of black hole. 
With $\nu_0=\nu_1=0$, we obtain $\det\mbox{M}=0 $; see \eqref{main1}.
Blue curves in Fig.~\ref{intersection1} (b) is the solution of $\det\mbox{M}=0 $ in \eqref{gen:3} for inside of black hole. 
%
\subsection{A  generic  working  case inside of the black hole:} 
\begin{figure}[!htb]
\centering 
{ \includegraphics[width=0.3\linewidth]{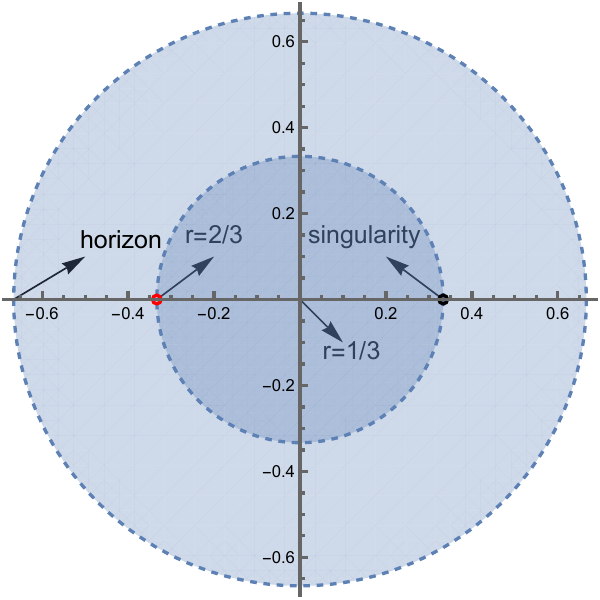}}  
\caption{ Domain of $y_{in}(\zeta_{a})$ at $a=1/3$.   } 
\label{delta}
\end{figure}
We change the coordinate $\zeta_a $ at $a=1/3$ in eq.(\ref{mm:2}).
And take $Y_{in}(\zeta_a) = \left( \frac{1}{3}-\zeta_a\right)^{\Delta -\frac{2 \lambda }{\sqrt{3}}}y_{in}(\zeta_a)$ into the new  eq.(\ref{mm:2}).    
Putting $a=1/3$ into \eqref{gen:1},
we have
\begin{equation} 
e_{n+2}=	A_n e_{n+1}+B_n e_n+ C_n e_{n-1}+ D_n e_{n-2}  \hspace{1.5cm}  \mbox{where} \; n \geq 2 	\label{zz:2}
\end{equation} 
with
\[\begin{cases} 
e_2=   A_0 e_1+B_0 e_0 \\
e_3= A_1 e_2+B_1 e_1+C_1 e_0.
\end{cases}\] 
where
\begin{equation}
\begin{cases}  
A_n= \frac{-20 \sqrt{3} \lambda +69 n+69}{26 n+52} \cr
B_n=  \frac{3 \left(-3 (\Delta -3) \Delta -8 \lambda ^2+9 \sqrt{3} \lambda +18 \sqrt{3} \lambda  n+18 n (n+1)\right)}{26 (n+1) (n+2)} \cr
C_n=  \frac{9 \left(3 (\Delta -3) \Delta +2 \left(\lambda ^2+3\right)-6 n^2+3 \sqrt{3} \lambda  n\right)}{13 (n+1) (n+2)} \cr
D_n = \frac{27 \left(-3 (\Delta -3) \Delta +4 \lambda ^2+2 \sqrt{3} \lambda -4 \sqrt{3} \lambda  n+3 (n-1) n-6\right)}{26 (n+1) (n+2)}
\end{cases}
\nonumber
\end{equation} 

From Birkhoﬀ–Adams theorem, we obtain
\begin{eqnarray} 
\lim_{n\to \infty} e_n  
&=&  \nu _0 \frac{  \left(-\frac{3}{2}\right)^n}{n}+\nu _1 \frac{  3^n}{n}+\nu _2  \frac{  2 \left(\frac{9}{13}\right)^{n/2}  \cos \left(n \tan ^{-1}\left(\frac{3 \sqrt{3}}{5}\right)\right)}{n^{\frac{2 \lambda }{\sqrt{3}}+1}}
\label{zz:3}
\end{eqnarray}

The biggest term in $\lim_{n\to \infty} e_n $ is  $  \frac{ 3^n}{n}.$ 
From ration test,
$$\lim_{n\to \infty} \left| \frac{e_{n+1}}{e_n}\right| |\zeta_{a}|=\lim_{n\to \infty} \left| \frac{ 3^{n+1}}{ n+1}\frac{n}{ 3^{n}}\right| |\zeta_{a}| = |3||\zeta_{a}|<1$$.

Therefore, $|\zeta_{a}|<1/3.$ 

The dominant behavior of $ y_{in}(\zeta_{a})$ is taken by
\begin{eqnarray} 
&&   y_{in}(\zeta_{a}) =\sum_{n=0 }^{\infty} e_n \zeta_{a}^n  \sim     \nu _0 \sum _{n }^{\infty } \frac{\left(-\frac{3}{2}\zeta_{a} \right)^n}{n}+ \nu _1 \sum _{n }^{\infty } \frac{( 3 \zeta_{a})^n}{n}   \nonumber
\end{eqnarray}

\begin{eqnarray} 
\mbox{When approaching   $\zeta_{a}= -\frac{1}{3}$:}  && \lim_{\zeta_{a} \to -\frac{1}{3}} y_{in}(\zeta_{a}) = \nu _0 \sum _{n }^{\infty } \frac{\left( \frac{1}{2} \right)^{n}}{n}+ \nu _1 \sum _{n }^{\infty } \frac{\left(-1\right)^{n}}{n} \nonumber
\end{eqnarray}    
\begin{eqnarray} 
\mbox{When approaching  $\zeta_{a}= \frac{1}{3}$ (singularity):}  
&&  \lim_{\zeta_{a} \to  \frac{1}{3}} y_{in}(\zeta_{a})= \nu _0 \sum _{n }^{\infty } \frac{\left( -\frac{1}{2} \right)^{n}}{n}+ \nu _1 \sum _{n }^{\infty } \frac{1}{n} \nonumber
\end{eqnarray}     

For well-definedness  at $\zeta_{a}= -\frac{1}{3} $ (because this is the regular point and it is located inside of a black hole) we require $\nu _1 =0$. As  $\nu _1 =0$, it implies to the regularity at the singular point.

Then, eq.(\ref{zz:3}) becomes 
\begin{eqnarray} 
\lim_{n\to \infty} e_n  
&=& \nu _0 \frac{  \left(-\frac{3}{2}\right)^n}{n}+	\nu _2  \frac{  2 \left(\frac{9}{13}\right)^{n/2}  \cos \left(n \tan ^{-1}\left(\frac{3 \sqrt{3}}{5}\right)\right)}{n^{\frac{2 \lambda }{\sqrt{3}}+1}}  
\end{eqnarray} 
The biggest term in $\lim_{n\to \infty} e_n $ is  $\frac{  \left(-\frac{3}{2}\right)^n}{n}.$ 
From ration test,
$$\lim_{n\to \infty} \left| \frac{e_{n+1}}{e_n}\right| |\zeta_{a}|=\lim_{n\to \infty} \left|\frac{  \left(-\frac{3}{2}\right)^{n+1}}{n+1} \frac{n}{ \left(-\frac{3}{2}\right)^{n}}\right| |\zeta_{a}| = |3/2||\zeta_{a}|<1.$$ 
Therefore, $|\zeta_{a}|<2/3.$ The radius of convergence is extended.

\begin{eqnarray} 
\mbox{When approaching  $\zeta_{a}= -\frac{2}{3}$ (horizon): }  && \lim_{\zeta_{a} \to -\frac{2}{3}} y_{in}(\zeta_{a}) = \nu _0 \lim_{\zeta_{a} \to -\frac{2}{3}}  \sum _{n }^{\infty } \frac{\left(-\frac{3}{2}\zeta_{a} \right)^n}{n} =\nu _0 \sum _{n }^{\infty } \frac{1}{n} \nonumber
\end{eqnarray} 
For regularity at the horizon, we require $\nu _0 =0$.

Therefore, as $\nu _0 =0$ and $\nu _1 =0$ are valid, the regularities of the singularity and the horizon are satisfied.

This means that 
\begin{footnotesize}
\be 
\nu _0=0, \; \nu _1=0  \leftrightarrow  \det \mbox{M}'  = 0
\label{rrr:61}
\ee
\end{footnotesize} 
See \eqref{main1}. Blue curves in Fig.~\ref{intersection1} (b) is the solution  of the $\det \mbox{M}'  = 0$ for inside of black hole.
\subsection{A generic   case with subtlety:} 
\begin{figure}[!htb]
\centering 
{ \includegraphics[width=0.3\linewidth]{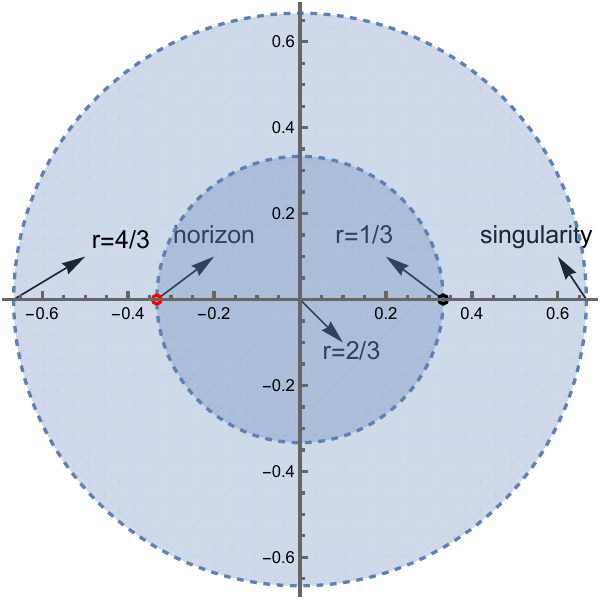}}  
\caption{ Domain of $y_{in}(\zeta_{a})$ at $a=2/3$.   } 
\label{delta}
\end{figure} 
We change the coordinate $\zeta_a $ at $a=2/3$ in eq.(\ref{mm:2}). 
And take $Y_{in}(\zeta_a) = \left( \frac{2}{3}-\zeta_a\right)^{\Delta -\frac{2 \lambda }{\sqrt{3}}}y_{in}(\zeta_a)$ into the new  eq.(\ref{mm:2}).    
Putting $a=2/3$ into \eqref{gen:1}, 
we have

%
\begin{equation} 
e_{n+2}=	A_n  e_{n+1}+B_n e_n+C_n e_{n-1}+D_n e_{n-2}  \hspace{1.5cm}  \mbox{where} \; n \geq 2 \nonumber  
\end{equation} 
with
\[\begin{cases} 
e_2=   A_0 e_1+B_0 e_0 \\
e_3= A_1 e_2+B_1 e_1+C_1 e_0.
\end{cases}\] 
where
\begin{equation}
\begin{cases}  
A_n= -\frac{28 \sqrt{3} \lambda +15 n+15}{38 (n+2)}  \cr
B_n=  \frac{3 \left(-12 (\Delta -3) \Delta -8 \lambda ^2-3 \sqrt{3} \lambda  (2 n+1)+72 n (n+1)\right)}{38 (n+1) (n+2)}  \cr
C_n=  \frac{9 \left(6 (\Delta -2) (\Delta -1)-2 \lambda ^2-12 n^2+9 \sqrt{3} \lambda  n\right)}{19 (n+1) (n+2)} \cr
D_n =\frac{27 \left(-3 (\Delta -3) \Delta +4 \lambda ^2+2 \sqrt{3} \lambda -4 \sqrt{3} \lambda  n+3 (n-1) n-6\right)}{38 (n+1) (n+2)}
\end{cases}
\nonumber
\end{equation} 

From Birkhoﬀ–Adams theorem, we obtain
\begin{eqnarray} 
\lim_{n\to \infty} e_n  
&=&  \nu_0 \frac{  (-3)^n}{n}+\nu_1 \frac{  \left(\frac{3}{2}\right)^n}{n}+	\nu_2 \frac{  2 \left(\frac{9}{19}\right)^{n/2}  \cos \left(n \tan ^{-1}\left(\frac{3 \sqrt{3}}{7}\right)\right)}{n^{\frac{2 \lambda }{\sqrt{3}}+1}}
\label{zz:1}
\end{eqnarray}  
The biggest term in $\lim_{n\to \infty} e_n $ is  $  \frac{  (-3)^n}{n}.$ 
From ration test,
$$\lim_{n\to \infty} \left| \frac{e_{n+1}}{e_n}\right| |\zeta_{a}|=\lim_{n\to \infty} \left| \frac{ (-3)^{n+1}}{ n+1}\frac{n}{ (-3)^{n}}\right| |\zeta_{a}| = |3||\zeta_{a}|<1$$.

Therefore, $|\zeta_{a}|<1/3.$  The dominant behavior of $ y_{in}(\zeta_{a})$ is taken by
\begin{eqnarray} 
&&   y_{in}(\zeta_{a}) =\sum_{n=0 }^{\infty} e_n \zeta_{a}^n  \sim    \nu_0 \sum _{n }^{\infty } \frac{(-3\zeta_{a})^n}{n}+ \nu_1 \sum _{n }^{\infty } \frac{( \frac{3}{2} \zeta_{a})^n}{n}   \nonumber
\end{eqnarray} 
\begin{eqnarray} 
\mbox{When approaching $\zeta_{a}=  -\frac{1}{3}$  (horizon):}  && \lim_{\zeta_{a} \to -\frac{1}{3}} y_{in}(\zeta_{a}) = \nu_0 \sum _{n }^{\infty } \frac{1}{n}+ \nu_1 \sum _{n }^{\infty } \frac{\left(-\frac{1}{2} \right)^{n}}{n}  \nonumber
\end{eqnarray}

For the horizon regularity we require $\nu_0 =0$. 
Then, eq.(\ref{zz:1}) becomes 
\begin{eqnarray} 
\lim_{n\to \infty} e_n  
&=& \nu_1 \frac{  \left(\frac{3}{2}\right)^n}{n}+	\nu_2 \frac{  2 \left(\frac{9}{19}\right)^{n/2}  \cos \left(n \tan ^{-1}\left(\frac{3 \sqrt{3}}{7}\right)\right)}{n^{\frac{2 \lambda }{\sqrt{3}}+1}}   
\end{eqnarray} 
The biggest term in $\lim_{n\to \infty} e_n $ is  $\frac{  \left(\frac{3}{2}\right)^n}{n}.$ 
From ration test,
$$\lim_{n\to \infty} \left| \frac{e_{n+1}}{e_n}\right| |\zeta_{a}|=\lim_{n\to \infty} \left|\frac{  \left(\frac{3}{2}\right)^{n+1}}{n+1} \frac{n}{ \left(\frac{3}{2}\right)^{n}}\right| |\zeta_{a}| = |3/2||\zeta_{a}|<1$$.

Therefore, $|\zeta_{a}|<2/3.$ The radius of convergence is extended.

\begin{eqnarray} 
\mbox{When approaching   $\zeta_{a}= -\frac{2}{3}$:}  && \lim_{\zeta_{a} \to -\frac{2}{3}}y_{in}(\zeta_{a}) = \nu_1 \lim_{\zeta_{a} \to -\frac{2}{3}} \sum _{n }^{\infty } \frac{( \frac{3}{2} \zeta_{a})^n}{n} = \nu_1 \sum _{n }^{\infty } \frac{(-1)^n}{n} \nonumber
\end{eqnarray}  
\begin{eqnarray} 
\mbox{When approaching  $\zeta_{a}=  \frac{2}{3}$  (singularity): }  && \lim_{\zeta_{a} \to \frac{2}{3}} y_{in}(\zeta_{a})= \nu_1 \lim_{\zeta_{a} \to  \frac{2}{3}} \sum _{n }^{\infty } \frac{( \frac{3}{2} \zeta_{a})^n}{n} = \nu_1 \sum _{n }^{\infty } \frac{1}{n} \nonumber
\end{eqnarray}  
For well-definedness  at  $\zeta_{a}= -\frac{2}{3}$, we require $\nu_1 =0$.   As consequence, we are able to admit the  regularity of the singularity.  With $\nu_0=0$,$\nu_1=0$, we obtain 
\be \lim_{n\to \infty} e_{n}=0.\nonumber\ee 
Therefore, 
\be 
\nu_0=0, \nu_1=0  \leftrightarrow \det \mbox{M}''  = 0.
\nonumber
\ee  
See \eqref{main1}. Blue curves in Fig.~\ref{intersection1} (b) are the solution  of $\det \mbox{M}''  = 0$ for inside of black hole.

\subsection{The horizon centered coordinate:}\label{horowitz2}  
we change the coordinate $\rho=1-\frac{1}{z}$  of $y_{in}(\rho)$ in eq.(\ref{mm:14}).  
\begin{figure}[!htb]
\centering
{ \includegraphics[width=0.3\linewidth]{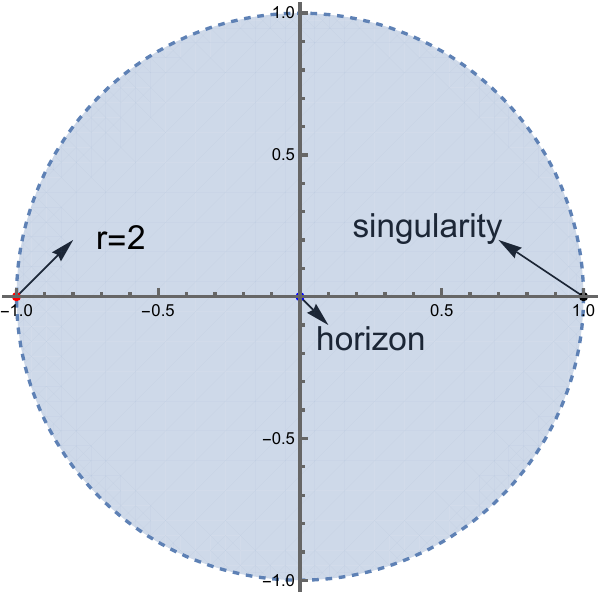}} 
\caption{ Domain of $y_{in}(\rho)$.   } 
\label{domain1}
\end{figure}  
At $\rho=0$ (horizon),  the differential equation has  only   one degenerate indicial root that develop log solution such as	  
\begin{equation}
y_{in}(\rho) = C  H_{1}(\rho) +D \left(  H_{1}(\rho) \log(\rho) +  H_{2}(\rho)  \right). \nonumber  
\end{equation} 
The first term is called as the first independent solution and the second term is called as the second independent one.
Set $D=0$ for the horizon regularity. 	 
Putting $ H_{1}(\rho) =\sum_{n=0}^{\infty} t_n \rho^n$ into a differential equation,  
we have
\begin{equation} 
t_{n+1}=\bar{A}_n t_n+\bar{B}_n t_{n-1}+\bar{C}_n t_{n-2} \nonumber  
\end{equation} 
$\hbox{  for  } n \geq 2$.   The first two  $t_{n}$'s are  $t_1=\bar{A}_0 t_0$ and  
$t_2=\bar{A}_1 t_1+\bar{B}_1 t_0$.
where
\begin{equation}
\begin{cases}  
\bar{A}_n=-\frac{(\Delta -3) \Delta +\sqrt{3} \lambda +2 n \left(\sqrt{3} \lambda -3 n-3\right)}{3 (n+1)^2}   \cr
\bar{B}_n= \frac{2 \left(3 (\Delta -3) \Delta -2 \lambda ^2-6 n^2+5 \sqrt{3} \lambda  n+6\right)}{9 (n+1)^2}  \cr
\bar{C}_n= \frac{-3 (\Delta -3) \Delta +4 \lambda ^2+2 \sqrt{3} \lambda -4 \sqrt{3} \lambda  n+3 (n-1) n-6}{9 (n+1)^2}  
\end{cases}
\nonumber
\end{equation}  
We have  \be 
\lim_{n\to \infty} t_n  = \beta_0 \left(\frac{1}{\sqrt{3}}\right)^n  \frac{     \cos \left(\frac{\pi  n}{6}\right)}{n^{\frac{2 \lambda }{\sqrt{3}}+1}}+\beta_1 \frac{1}{n}. \label{mm:205}\ee

The biggest term in $\lim_{n\to \infty} t_n $ is  $  \frac{ 1}{n}.$ 
From ration test, we obtain $|\rho|<1.$ 

At $r=2$,  \eqref{mm:205} gives us the asymptotic series for $H_{1}(\rho)$ such as $\sum_{n }^{\infty} t_n \rho^n  \sim  \beta_1 \sum _{n }^{\infty } \frac{(-1)^n}{n} $. Similarly, at the singularity,  $\sum_{n }^{\infty} t_n \rho^n  \sim   \beta_1 \sum _{n }^{\infty } \frac{1}{n} $.
And we allow the well-definedness at ordinary point $r=2$ by setting  $\beta_1=0$ and it simultaneously implies regularity at the sigularity.   
In Fig.~\ref{intersection1} (b), the blue curve is obtained by resolving the singularity at the horizon in a coordinate system with the center of the black hole as the origin or in a coordinate system with any ordinary point inside the black hole at $1 \leq z \leq  \infty$ as the origin. This blue curve is also obtained by applying $\lim_{n \to \infty} (n+1)t_{n+1} =0$.

In conclusion, if we set $\zeta_a=a-\frac{1}{z}$  of $Y_{in}(\xi)$ in eq.(\ref{mm:14}) at $0 \leq a \leq 1$,   the horizon regularity condition with well-definedness implies the regularity at the center of a black hole.

\section{Stability of log divergence under the  backreaction:}\label{yoon}	
Full solution near singularity with the backreaction on the
spacetime geometry is given by \cite{Horo2009,Horo2011,Hartnoll:2020fhc}

\begin{small}
\bea
&& \Psi ''(\xi )+   \left(\frac{f'(\xi )}{f(\xi )}-\frac{\chi '(\xi )}{2}+\frac{4}{\xi }\right)  \Psi '(\xi ) +  \left(\frac{q^2 \Phi (\xi )^2 e^{\chi (\xi )}}{\xi ^4 f(\xi )^2}-\frac{m^2  }{\xi ^2 f(\xi )}\right) \Psi (\xi ) =0, \label{dis:13} \\
&& \Phi ''(\xi )+ \left(\frac{\chi '(\xi )}{2}+\frac{2}{\xi }\right)\Phi '(\xi )   -\frac{ 2 q^2 \Psi (\xi )^2}{\xi ^2 f(\xi )}\Phi (\xi ) =0  \label{dis:14} \\
&&  \chi '(\xi )+\xi  \Psi '(\xi )^2 + \frac{q^2 \Phi (\xi )^2  \Psi (\xi )^2 e^{\chi (\xi )} }{\xi ^3 f(\xi )^2}=0  \label{dis:15}\\
&&   f'(\xi )+\frac{\xi  \Psi '(\xi )^2}{2} f(\xi )  +\frac{3 (f(\xi )-1)}{\xi }+\frac{q^2 \Phi (\xi )^2   \Psi (\xi )^2 e^{\chi (\xi )}}{2 \xi ^3 f(\xi )}+\frac{m^2 \Psi (\xi )^2}{2 \xi }+\frac{  \Phi '(\xi )^2 e^{\chi (\xi )}}{4 \xi }=0, 
\label{dis:16}
\eea
\end{small}   
here,  $\xi=1/z$.  For $\xi  \rightarrow 0$, 
\begin{small}
\bea
&& f=-f_K \xi^{-3-\alpha^2} + \cdots, \label{dis:17} \\
&& \Psi = \alpha \sqrt{2} \log (1/\xi)+ \cdots, \label{dis:18} \\
&& \chi =2 \alpha^2 \log (1/\xi) +\chi_k +\cdots,\label{dis:19}
\eea
\end{small}
Here, $m^2  =\Delta (\Delta-3)$.  

Non-linear second order differential equation of $\Psi(\xi)$ has regular singular points at horizon and singularity (the center of a black hole). As the method of Frobenius is applied near the horizon and singularity, each has one degenerate indicial root such as zero. Therefore, a log solution should be appeared at each singular point.

Non-linear second order differential equation of $\Psi(\xi)$ near $\xi =0$ must have two independent solution such that 
\bea \Psi (\xi )=\mathcal{A} S _1(\xi )-\alpha \sqrt{2} \left( S _1(\xi ) \log (\xi )+  S _2(\xi )\right) \label{dis:20}\eea

at $|\xi|<1$. Here, $\Psi_1(\xi)$ is named as the first kind of a solution and $\Psi_2(\xi)$ is called as the second kind of a solution. 
$\Psi_1(\xi)$ and $\Psi_2(\xi) $ are satisfied with following two differential equation such that
\begin{scriptsize} 
\bea
&& S _1''(\xi )+   \left(\frac{f'(\xi )}{f(\xi )}-\frac{\chi '(\xi )}{2}+\frac{4}{\xi }\right) S _1'(\xi ) +  \left(\frac{q^2  \Phi (\xi )^2 e^{\chi (\xi )}}{\xi ^4 f(\xi )^2}-\frac{m^2}{\xi ^2 f(\xi )}\right) S _1(\xi ) =0,  \label{dis:22}\\
&& S _2''(\xi ) + \left(\frac{f'(\xi )}{f(\xi )}-\frac{\chi '(\xi )}{2}+\frac{4}{\xi }\right)  S _2'(\xi )  +  \left(\frac{q^2   \Phi (\xi )^2 e^{\chi (\xi )}}{\xi ^4 f(\xi )^2}-\frac{m^2}{\xi ^2 f(\xi )}\right) S _2(\xi ) =-   \left(\frac{f'(\xi )}{\xi  f(\xi )}+\frac{3}{\xi ^2}-\frac{\chi '(\xi )}{2 \xi }\right)  S _1(\xi ) -\frac{2  }{\xi }S _1'(\xi ) .\label{dis:23}  
\eea
\end{scriptsize}
Asymptotic solution of $\Phi(\xi)$ becomes
\bea
\lim_{\xi \to  0}\Phi(\xi) = V + W\; \xi^{\alpha^2 -1} \label{dis:24}
\eea
Near $\xi=1$, we can expand $f(\xi)=f_0 (1-\xi)+ f_1 (1-\xi)^2+\cdots $ at $f_0<0$.
And we have $\chi(1), \chi'(1) < \infty$ and $\Phi(1)=0$. With these conditions, eq.(\ref{dis:13}) can be expressed as 
\bea
\frac{d^2 \Psi }{d \xi^2} + \frac{1}{\xi-1} \frac{d \Psi }{d \xi} - \frac{\Delta (\Delta-3)}{f_0 (1-\xi)}  \Psi =0  \label{dis:25} 
\eea
near $\xi=1$. The asymptotic solution of eq.(\ref{dis:25}) is $ \lim_{\xi \to  1} \Psi(\xi) \sim \log(1-\xi)$. Then,  we have the following asymptotic expression in eq.(\ref{dis:20}); 
\bea
&&  \lim_{\xi \to  1} S_1(\xi) = a_1 \log(1-\xi) , \label{dis:26}   \\
&& \lim_{\xi \to  1} S_2(\xi) = a_2 \log(1-\xi). \label{dis:27} 
\eea
Here,  $a_1$ and $a_2$ consists of $\Delta$, $q$,  $\rho$ and $\left< \mathcal{O}_{\Delta}\right>$.

And
\bea  &&  \lim_{\xi \to 1} \Psi(\xi)  =\lim_{\xi \to 1}  \left(\mathcal{A} S _1(\xi )-\alpha \sqrt{2} \left( S _1(\xi ) \log (\xi )+  S _2(\xi )\right) \right)   = \lim_{\xi \to 1}  \left( \mathcal{A} S_{1}(\xi) -\alpha \sqrt{2}  S_{2}(\xi) \right) \nonumber \\
&& \sim  \left(  \mathcal{A} a_1 -\alpha \sqrt{2}   a_2 \right) \log(1-\xi)\eea 
According to the horizon regulation condition, we must set $ \alpha=  \frac{ \mathcal{A} a_1 }{\sqrt{2} a_2  }$, $\log(1-\xi)$ behavior does not exist any more. This process makes that  $\mathcal{A} S _1(\xi )-\alpha \sqrt{2}  \left( S _1(\xi ) \log (\xi )+  S _2(\xi )\right) $ converge at the horizon. However, if we take the derivative of $ \Psi(\xi) $, it is not convergent any more. Because  
\begin{eqnarray} 
&& \lim_{\xi \to 1}\frac{d\Psi(\xi) }{d \xi}= \lim_{\xi \to 1}\frac{d }{d \xi} \Big(\mathcal{A} S _1(\xi )-\alpha \sqrt{2}  \left( S _1(\xi ) \log (\xi )+  S _2(\xi )\right)\Big)  \nonumber\\
&=& -\alpha \sqrt{2} \lim_{\xi \to 1} \frac{d}{d \xi} \Big(   S_{1}(\xi) \log \xi    \Big)+ \lim_{\xi \to 1} \frac{d }{d \xi} \Big( \mathcal{A} S_{1}(\xi) -\alpha \sqrt{2}   S_{2}(\xi)  \Big) \nonumber\\
&&\sim -\alpha \sqrt{2}  \lim_{\xi \to 1} \frac{d }{d \xi} \Big(  S_{1}(\xi) \log \xi  \Big) \rightarrow \infty.
\end{eqnarray} 
Notice that  we make $\mathcal{A} S_{1}(\xi) -\alpha \sqrt{2}  S_{2}(\xi)  $ convergent at $\xi=1$ on the above by setting $ \alpha=  \frac{ \mathcal{A} a_1 }{\sqrt{2} a_2  }$. In other words, it is analytic at the horizon and  differentiable at that point. So $$\lim_{\xi \to 1} \frac{d }{d \xi} \Big(\mathcal{A} S_{1}(\xi)  -\alpha \sqrt{2}    S_{2}(\xi)   \Big) < \infty.$$
Therefore, we should remove $\log \xi$ term at $\xi=\infty$ by setting $\alpha =0$. To make horizon regularity, $a_1=0$ should be also satisfied.

We reach the conclusion such that  $\alpha =0$.	
Then, the series expansion of the vector potential near $\xi=0$ becomes
\bea 
\Phi (\xi)= V \phi_0 (\xi)+ W \xi^{-1}  \phi_1 (\xi).  \label{dis:32}   
\eea

We should set $W=0$ (If   $W \ne 0$, the method of Frobenius is failed in eq.(\ref{dis:16}) near $\xi =0$ ). From these calculations,  we obtain interesting results such as
$$  \lim_{z \rightarrow \infty}\psi (z), \lim_{z \rightarrow \infty}\Phi (z), \lim_{z \rightarrow \infty}\chi(z) \rightarrow \mbox{constant}. $$ 

We have three different constrained conditions such as (1) horizon regularity condition for scalar field,  (2) the regularity at the center of a black hole for scalar field (3) the regularity at the center of a black hole for vector potential. And we have 4 unknown parameters such as $\Delta$, $q$,  $\rho$ and $\left< \mathcal{O}_{\Delta}\right>$. If one parameter is given, other parameters are determined automatically from 3 constrained conditions.

\section{ Near critical temperature  for 3+1 dimensional system:
}\label{hahaha}

At  the critical temperature $T_c$, $\Psi =0$, so the field equation   eq.(\ref{qqq:1})  for the electrostatic potential reduces to  $\Phi^{''}=0$. Then, we may set \cite{Siop2010} 
\begin{equation}
\Phi(z)= \tilde{\lambda}  r_c (1-x) \hspace{1cm}\mbox{where}\;\;\tilde{\lambda} =\frac{\rho}{r_c^3}
\label{qq:9}
\end{equation}
where $x=z^2$ and $r_{c}$ is horizon radius at the critical temperature. For $T\rightarrow T_c$, the field equation $\Psi$ approaches to \cite{Choun:2021pvs}
\begin{equation}
-\frac{d^2 \Psi }{d x^2} +\frac{1+x^2}{x(1-x^2)}\frac{d \Psi }{d x}+  \frac{m^2}{4 x^2(1-x^2)} \Psi =\frac{\lambda ^2}{4x(1+x)^2}\Psi
\label{qq:10}
\end{equation}
where $\lambda = g\tilde{\lambda}$. The critical temperature is given by  \cite{Siop2010,Choun:2021pvs}
\be
T_c 
=\frac{d}{4\pi}r_c
=\frac{1}{ \pi} \left(\frac{g \rho }{\lambda }\right)^{\frac{1}{3}},
\label{si:1}
\ee
for $d=4$. 
Factoring out the behavior near  	$x=0$ and 
$x=-1$, we  have 
\begin{equation}
\Psi(x)= \frac{\left< \mathcal{O}_{\Delta}\right>}{\sqrt{2}r_h^{\Delta}}x^{\frac{\Delta}{2}} F(x)  \;\;\mbox{where}\;\;F(x)=
(1+x)^{-\lambda /2}Y_{out}(x).
\label{qq:11}
\end{equation}   
Then, $F$ is normalized as $F(0)=1$  and  we obtain
\begin{small}
\begin{eqnarray}
&&\frac{d^2 Y_{out} }{d x^2} + \left( \frac{\rho_0}{x}+\frac{\rho_1}{x-1}+\frac{\rho_2 }{x+1}\right) \frac{dY_{out}}{d x}  \nonumber\\
&&+\left(\frac{w_0}{x}+\frac{w_1}{x-1}+\frac{w_2}{x+1}\right)Y_{out}=0,
\label{qq:12}
\end{eqnarray}
\end{small}
\begin{equation}
\hbox{where }		\begin{cases} \rho_0=\Delta-1, \quad 
\rho_1=1 , \quad 
\rho_2=1-\lambda ,\cr 
w_0=\frac{\lambda }{2}\left(-\Delta +\frac{\lambda }{2}+1\right), \cr
w_1=\frac{1}{8}(\Delta ^2-2 \lambda ), \cr
w_2=\frac{1}{8} \left(-\Delta ^2+4 \Delta  \lambda -2 \lambda ^2-2 \lambda \right).        
\end{cases}
\nonumber
\end{equation} 
Eq.(\ref{qq:12})  has four regular singular points at $x=0,1,-1,\infty$, called the Heun's differential equation \cite{Ronv1995}. 
A  three-term recurrence relation starts appear by putting $Y_{out}(x)= \sum_{n=0}^{\infty } d_n x^{n}$ at $|x|<1$ into (\ref{qq:12}):
\begin{equation}
d_{n+1} = A_n \;d_n + B_n \;d_{n-1}    \quad
\label{qq:13}
\end{equation}
for   $n \geq 1$, with
\begin{equation} 
A_n=  \frac{\frac{\lambda }{2}\left(2n+\Delta-1-\frac{\lambda }{2}\right)}{(n+1)(n+ \Delta-1)}    \quad
B_n= \frac{\left( n -1+\frac{\Delta}{2}-\frac{\lambda }{2}\right)^2}{(n+1)(n+ \Delta-1)}.   
\label{qq:14}
\end{equation}
The first two  $d_{n}$'s are determined by   $  d_1= A_0 d_0 $ and  $d_{0}=1$.  
Eq.(\ref{qq:11}), Eq.(\ref{qq:13}) and Eq.(\ref{qq:14})  give us the  boundary condition $
F^{'}(0)=0.$


\subsection{ Scaling dimension from the black hole interior: Method (I) }\label{Tcc}  
We can determine the discrete values of allowed scaling dimension, if we include  the interior of the black hole as well as outside as our solution  of  the Heun's equation,  
eq.(\ref{qq:12}) must be a polynomial, because  the infinite series in eq.(\ref{qq:12}) is only valid at $x\leq 1$: If we include Pincherle theorem, we of course  have a convergent series solution at the horizon.  If the degree of the polynomial is $N$, then 
we need to impose  
\be
B_{N+1}=d_{N+1}=0 \quad  \mbox{ for degree  } N \in \mathbb{N}_{0},  \label{aa:1}
\ee   
In our case, two parameters  $\lambda$ and $\Delta$ should be quantized, because  the solutions  correponds to the intersection of two $N+1^{\mbox{th}}$  polynomials defined by eqs.(\ref{aa:1}). From eq.(\ref{qq:14}), 
$B_{N+1}=0$ gives 
\be
\lambda =2N+\Delta. \label{Omega}
\ee  

Also, $d_{N+1}=0$  gives  a $N+1$-th order polynomial  in $\Delta$ that we denote it as $ {\cal P}_{N+1} =0$. 
These  	tell us that 
\begin{description} 
\item[$N=0$]  
$ (\Delta,\lambda)=(2,2),   $
\item[$N=1$ ]  
$ (\Delta,\lambda)=(3.635,5.635),  $  
\item[$N=2$ ]  
$ (\Delta,\lambda)=(2,2),\; (5.05,9.05),  $  
\item[$N=3$]  
$ (\Delta,\lambda)=(3.716,9.716),\; (6.436,12.436)  $
\item[$N=4$ ]  
$ (\Delta,\lambda)=(2,2),\; (5.2,13.2),\; (7.83,15.83)  $  
\item[$N=5$ ]  
$ (\Delta,\lambda)=  (3.76,13.76),\;  (6.61,16.61),\; (9.23,19.23).  $ 
\end{description}  
The relation \eqref{Omega} tells us that  solotions with lower $\Delta$ and $N$ are more stable under the perturbation since they give lower eigenvalue $\lambda$.   
\vskip.2cm\subsection{ Scaling dimension from the black hole interior: Method (II) }\label{AA} 	 
In the previous section, we show how to compute scaling dimension $\Delta$ by considering inside and outside of the black hole. In this section,  we reproduce the same result with different method.		
In terms of methodology, we will divide the solutions of the equations into two sections, outside ($0<x\leq 1$) and inside  ($x\geq 1$) the black hole, and then gently connect the solutions of these two equations at the event horizon at $x=1$: In general, the two solutions can be defined to be smoothly connected on the horizon by adjusting the parameter values that do not converge on the horizon.

For sufficiently large $n$ in eq.(\ref{qq:13}), $d_{n}  $ has two linearly independent solutions $d_{n}^{(1)}$ and $d_{n}^{(2)}$. It can be described as  $\lim_{n \to \infty} d_{n} = a \; d_{n}^{(1)} +b\; d_{n}^{(2)}$. Here, $a,b$ consists of $(\Delta, \lambda)$.  

One can show that  \cite{Jone1980} for sufficiently large $n$,  
\begin{equation}
\begin{cases}  d_{n}^{(1)}\sim  (-1)^n  n^{-1-\lambda }  , \cr
d_{n}^{(2)}\sim n^{-1} , 
\end{cases}
\label{qq:32}
\end{equation}
at $\lambda >0$.


A harmonic series   $\sum_{n}^{\infty}d_n^{(2)} x^n$ develops $b \log(1-x)$ behavior near horizon. 	 
For the horizon regularity, we require $b=0$. 
With this condition, it makes convergent at the horizon.   

\begin{figure}[!htb]
\centering
{ \includegraphics[width=0.25\linewidth]{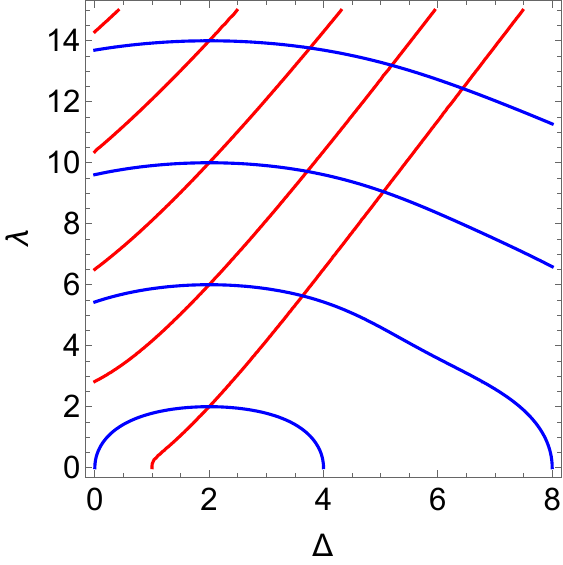}} 
\caption{  $\lambda $  vs $\Delta$ } 
\label{intersection}
\end{figure}

$b=0$ is equivalent to 
\begin{equation}
\lim_{n\to \infty}  (n+1) d_{n+1}=0  
\label{discrete}
\end{equation}   
As a result we  will  have a relation between  the scaling dimension and coupling.    In Fig.~\ref{intersection},  red curves are solutions of the transcendental equation  (\ref{discrete})  of $\lambda$ with given $\Delta$ for outside of balck hole. 
This algorithm for finding $\lambda$ with given $\Delta$ is only valid at $0<x \leq 1$. Now we consider inside of a black hole at $x\geq 1$.

First, we change a coordinate $\xi= 1/x$ in  eq.(\ref{qq:12}). And factoring out the behavior near $\xi =0$, we have
\begin{equation} 
Y_{in}(\xi)= \mathcal{N}\xi^{\gamma} G(\xi). 
\label{eee:1}
\end{equation}
here, $\mathcal{N}$ is a connection coefficient.
The differential equation has  only   one degenerate indicial root, and, the second independent solution having $\log(\xi)$ term should be dropped for the horizon regularity; see the section \ref{gaga}. Because, otherwise, higher derivative at the horizon does not converge.
Later, we decide this value by letting $Y_{out}(1)=Y_{in}(1)$; notice that $Y_{out}'(1)=Y_{in}'(1)$ valid automatically if $Y_{out}(1)=Y_{in}(1)$ is satisfied; because $Y_{in}(\xi)$ has only one indicial root  $\gamma=\gamma_1=\frac{\Delta -\lambda }{2}$.    
Once again,   $Y_{in}(\xi)$ has only one independent solution rather than two solutions for this case.
And we obtain  
\begin{eqnarray}
\frac{d^2 G}{d \xi^2} + \left( \frac{\theta_0}{\xi}+\frac{\theta_1}{\xi-1}+\frac{\theta_2 }{\xi+1}\right) \frac{d G}{d \xi}   +\left(\frac{\nu_0}{\xi}+\frac{\nu_1}{\xi-1}+\frac{\nu_2}{\xi+1}\right) G =0,
\label{eee:2}
\end{eqnarray} 
\begin{equation}
\hbox{where }		\begin{cases} 
\theta_0= 1, \quad 
\theta_1= 1, \quad 
\theta_2= 1-\lambda,\cr 
\nu_0= \frac{1}{4} (\lambda -2) \lambda, \cr
\nu_1= \frac{1}{8} \left(-\Delta ^2+4 \Delta -2 \lambda \right), \cr
\nu_2= \frac{1}{8} \left(\Delta ^2-4 \Delta -2 \lambda ^2+6 \lambda \right).        
\end{cases} 
\nonumber
\end{equation} 
Eq.(\ref{eee:2}) is also the Heun's differential equation \cite{Ronv1995}. 
A  three-term recurrence relation starts appear by putting $F(\xi)= \sum_{n=0}^{\infty } c_n \xi^{n}$ at $|\xi|<1$ into (\ref{eee:2}): 
\begin{equation}
c_{n+1}= \widetilde{A}_n \;c_n + \widetilde{B}_n \;c_{n-1},   \quad
\label{eee:3}
\end{equation}
for   $n \geq 1$, with
\begin{equation} 
\widetilde{A}_n = \frac{\lambda  n -\frac{1}{4} (\lambda -2) \lambda  }{(n+1)^2} \quad
\widetilde{B}_n =  \frac{\frac{1}{4} (\lambda -\Delta ) (\Delta +\lambda -4)+(n-1) (-\lambda +n+1)}{(n+1)^2}.   
\label{eee:4}
\end{equation}
The first two  $c_{n}$'s are determined by   $  c_1= \widetilde{A}_0 c_0$ and  $c_{0}=1$. 
We have $\lim_{n\to  \infty}c_{n} = \widetilde{a} \; c_{n}^{(1)} +\widetilde{b}\; c_{n}^{(2)}$. Here, $\widetilde{a}, \widetilde{b}$ consists of $\Delta$ and $\lambda$. $ c_{n}^{(i)}$ is an independent solution of $c_n$.  

From eq.(\ref{eee:8}) and eq.(\ref{eee:9}),  two asymptotic expansions of $c_n$ in  eq.(\ref{eee:3}) are
\begin{equation}
\begin{cases}  c_{n}^{(1)}\sim  (-1)^n  n^{-1-\lambda }  , \cr
c_{n}^{(2)}\sim n^{-1} , 
\end{cases}
\label{eee:10}
\end{equation}
at $\lambda >0$.


A harmonic series   $\sum_{n}^{\infty}c_n^{(2)} x^n$ develops $\widetilde{b} \log(1-x)$ behavior near horizon. 	 
For the horizon regularity, we require $\widetilde{b}=0$.  
With this condition,   it makes convergent at the horizon. 
$\widetilde{b}=0$ is equivalent to 
\begin{equation}
\lim_{n\to \infty}  (n+1) c_{n+1}=0  
\label{discrete1}
\end{equation}      
As a result we  will  have a relation between  the scaling dimension and coupling.    In Fig.~\ref{intersection},  blue curves are solutions of the transcendental equation  (\ref{discrete1})   of $\lambda$ with given $\Delta$ for inside of balck hole. Fig.~\ref{intersection} shows us that every intersection points between red and blue curves are equivalent  to $(\Delta, \lambda)$'s obtained by polynomial solutions; refer to the values of $(\Delta, \lambda)$ listed under eq.(\ref{Omega}).

\end{document}